\documentclass[preprint,3p,12pt]{elsarticle}

\usepackage{amssymb}
\usepackage{amsmath}
\usepackage{booktabs}
\usepackage{nicefrac}
\usepackage{graphicx}
\usepackage{lscape}
\usepackage{multicol}
\usepackage{tabularray}
\usepackage{subcaption}
\usepackage{nicematrix}
\usepackage{gensymb}
\usepackage{makecell}
\usepackage{hyperref}

\usepackage{lineno}

\biboptions{sort&compress}

\journal{Composites Part A}

\begin{document}
\newcommand{\I}{\rm{I}}
\newcommand{\II}{\rm{II}}
\newcommand{\lce}{l}
\newcommand{\beam}{\rm{beam}}
\newcommand{\btm}{\rm{bot}} 
\newcommand{\CE}{\rm{CE}}
\newcommand{\CR}{\rm{CR}}
\newcommand{\dd}{\mathrm{d}} 
\newcommand{\external}{\rm{ext}}
\newcommand{\h}{\rm{h}}
\newcommand{\internal}{\rm{int}}
\newcommand{\len}{\rm{\mathcal{l}}}
\newcommand{\truss}{\rm{truss}}
\newcommand{\tp}{\rm{top}} 
\newcommand{\transpose}{^{\rm{T}}}
\newcommand{\invtranspose}{^{\rm{-T}}}
\newcommand{\xl}{{\hat{x}}}
\newcommand{\yl}{{\hat{y}}}
\newcommand{\inv}{\makebox[0pt][l]{$^{-1}$}}
\newcommand{\DeltaI}{\Delta_{\text{I}}}
\newcommand{\DeltaII}{\Delta_{\text{II}}}
\newcommand{\DeltaIII}{\Delta_{\text{III}}}
\newcommand{\wtopce}{w^{\text{topCE}}}
\newcommand{\wbotce}{w^{\text{botCE}}}
\newcommand{\utopce}{u^{\text{topCE}}}
\newcommand{\ubotce}{u^{\text{botCE}}}
\newcommand{\vtopce}{v^{\text{topCE}}}
\newcommand{\vbotce}{v^{\text{botCE}}}
\newcommand{\wtop}{w^{\text{top}}}
\newcommand{\wbot}{w^{\text{bot}}}
\newcommand{\utop}{u^{\text{top}}}
\newcommand{\ubot}{u^{\text{bot}}}
\newcommand{\vttop}{v^{\text{top}}}
\newcommand{\vbot}{v^{\text{bot}}}
\newcommand{\htop}{h^{\text{top}}}
\newcommand{\hbot}{h^{\text{bot}}}
\newcommand{\thetatop}{\theta^{\text{top}}}
\newcommand{\thetabot}{\theta^{\text{bot}}}

\newcommand{\ba}{\boldsymbol{\mathrm{a}}}
\newcommand{\bA}{\boldsymbol{\mathrm{A}}}
\newcommand{\bb}{\boldsymbol{\mathrm{b}}}
\newcommand{\bB}{\boldsymbol{\mathrm{B}}}
\newcommand{\bBN}{\bB_{\text{N}}}
\newcommand{\bC}{\boldsymbol{\mathrm{C}}}
\newcommand{\D}{\boldsymbol{\mathrm{D}}}
\newcommand{\Dce}{\D^{\CE}}
\newcommand{\bE}{\boldsymbol{\mathrm{E}}}
\newcommand{\bF}{\boldsymbol{\mathrm{F}}}
\newcommand{\f}{\boldsymbol{\mathrm{f}}} 
\newcommand{\fext}{\f_{\external}}
\newcommand{\fint}{\f_{\internal}}
\newcommand{\fhatext}{\hat{\f}_{\external}}
\newcommand{\bH}{\boldsymbol{\mathrm{H}}}
\newcommand{\bI}{\boldsymbol{\mathrm{I}}}
\newcommand{\bJ}{\boldsymbol{\mathrm{J}}}
\newcommand{\K}{\boldsymbol{\mathrm{K}}}
\newcommand{\Kmat}{\K_{\mathrm{mat}}}
\newcommand{\Kgeo}{\K_{\mathrm{geo}}}
\newcommand{\bM}{\boldsymbol{\mathrm{M}}}
\newcommand{\bn}{\boldsymbol{\mathrm{n}}}
\newcommand{\bN}{\boldsymbol{\mathrm{N}}}
\newcommand{\bp}{\boldsymbol{\mathrm{p}}}
\newcommand{\bq}{\boldsymbol{\mathrm{q}}}
\newcommand{\bqce}{\bq^{\CE}}
\newcommand{\bQ}{\boldsymbol{\mathrm{Q}}}
\newcommand{\bU}{\boldsymbol{\mathrm{U}}}
\newcommand{\bR}{\boldsymbol{\mathrm{R}}}
\newcommand{\bS}{\boldsymbol{\mathrm{S}}}
\newcommand{\bt}{\boldsymbol{\mathrm{t}}}
\newcommand{\bT}{\boldsymbol{\mathrm{T}}}
\newcommand{\bu}{\boldsymbol{\mathrm{u}}}
\newcommand{\bv}{\boldsymbol{\mathrm{v}}}
\newcommand{\bw}{\boldsymbol{\mathrm{w}}}
\newcommand{\bW}{\boldsymbol{\mathrm{W}}}
\newcommand{\bX}{\boldsymbol{\mathrm{X}}}
\newcommand{\bx}{\boldsymbol{\mathrm{x}}}

\newcommand{\balpha}{\boldsymbol{\mathrm{\alpha}}}
\newcommand{\bDelta}{\boldsymbol{\mathrm{\Delta}}}
\newcommand{\beps}{\boldsymbol{\mathrm{\epsilon}}}
\newcommand{\bgamma}{\boldsymbol{\mathrm{\gamma}}}
\newcommand{\bPhi}{\boldsymbol{\Phi}}
\newcommand{\bsig}{\boldsymbol{\mathrm{\sigma}}}
\newcommand{\btau}{\boldsymbol{\mathrm{\tau}}}
\newcommand{\bxi}{\boldsymbol{\mathrm{\xi}}}

\newcommand{\minus}{\scalebox{0.75}[1.0]{$-$}}
\begin{frontmatter}



\title{A new approach for the determination of through-thickness and free-edge stresses in composite laminates based on structural elements}


\author[inst1]{Xiaopeng Ai} 
\author[inst1]{Christos Kassapoglou}
\author[inst2,inst1]{Boyang Chen\corref{cor1}}
 \ead{boyangchen@suda.edu.cn}
 \cortext[cor1]{Corresponding author}

\affiliation[inst1]{organization={Department of Aerospace Structures and Materials, Faculty of Aerospace Engineering, Delft University of Technology},
            addressline={Kluyverweg 1}, 
            city={Delft},
            postcode={2629 HS}, 
            country={The Netherlands}}

\affiliation[inst2]{organization={School of Optoelectronic Science and Engineering, Soochow University},
            addressline={Shizi Street 1}, 
            city={Suzhou},
            postcode={215031}, 
            country={China}}

\begin{abstract}
Thin shell elements based on the Kirchhoff-Love hypothesis account for only three stress components: the in-plane normal stresses ($\sigma_x$ and $\sigma_y$) and the in-plane shear stress ($\tau_{xy}$). The out-of-plane stress components required by conventional three-dimensional (3D) damage criteria are unavailable. As a result, damage initiation in composite laminates cannot be accurately predicted. This paper presents a method for capturing free-edge effects in multilayer structural elements with arbitrary composite lay-ups based on the Kirchhoff–Love hypothesis. The proposed formulation employs structural cohesive elements to model composite laminates and utilizes an accurate penalty stiffness derived from the resin-rich layers of the laminate. By accurately characterizing the interfacial mechanical response, out-of-plane stress components can be recovered. 
\end{abstract}

\begin{graphicalabstract}
\end{graphicalabstract}

\begin{highlights}
\item Research highlight 1
\item Research highlight 2
\end{highlights}

\begin{keyword}
Composites \sep Free-edge effect \sep Cohesive element \sep Interlaminar stresses


\end{keyword}

\end{frontmatter}



\section{Introduction}
\label{sec:Introduction}
Fiber-reinforced composites are popular choices for lightweight structures in aerospace \cite{krueger2015summary,kaddour2013mechanical}. The prediction of strength and damage tolerance is an indispensable aspect in the design of composite aircraft structures \cite{kassapoglou2010design}. The shell element is widely used in the simulation of thin-walled composite structures due to its efficiency \cite{carrera2017shell,soltani2021interlaminar}. Since the traditional thin-shell element can only obtain in-plane stress, important information needed for out-of-plane failure analysis and damage prediction is missing. 

In the past, many researchers conducted a large amount of research in this field. Most
authors solve the shear stress along the thickness direction based on the thick shell theory
of the Mindlin-Reissner hypothesis~\cite{reddy1993evaluation,groh2016computationally,szekrenyes2014stress}. The out-of-plane normal stress is then obtained through the stress recovery method \cite{leong2023adaptive,LEONG2024107974}. However, these methods only consider the in-plane failure mechanisms, such as matrix cracking and fiber failure, while neglecting interlaminar delamination. When considering delamination, cohesive elements (CEs) are widely used as the interface element based on the cohesive zone model (CZM) \cite{barenblatt1962mathematical,dugdale1960yielding}. However, when Mindlin–Reissner shell elements are used, compatible cohesive elements still face the bottleneck of the cohesive zone limitation~\cite{russo2020overcoming,balducci2024overcoming}. In previous studies, the authors overcame the limitation of conventional cohesive zone modeling by developing high-order shell elements and compatible cohesive elements. The proposed formulation was validated through delamination simulations of a series of benchmark problems~\cite{ai2025structural}. However, the shell element employed in the proposed formulation is based on the Kirchhoff-Love hypothesis. The previous approaches for recovering out-of-plane stresses developed for Mindlin–Reissner shell elements are not applicable. A dedicated approach is required to recover the out-of-plane stress field in the Kirchhoff–Love shell element.

\citet{framby2016assessment,framby2017adaptive} proposed a stress recovery method based on the Kirchhoff–Love hypothesis. The method recovers the out-of-plane stresses from the in-plane stress field of shell elements. It accurately predicts the through-thickness stress distribution in thin-shell structures. However, it cannot simulate delamination initiation and propagation because it does not incorporate cohesive elements.
\citet{LEONG2024107974} adopted this method by incorporating cohesive elements for damage analysis. His approach only recovered the out-of-plane normal stress, while the out-of-plane shear stresses were obtained from the Mindlin–Reissner plate theory. In addition, the shell elements adopted in Fr{\"a}mby's method are not high-order shell elements. When combined with cohesive elements, the mesh-density limitation imposed by the cohesive zone model still remains. By applying the stress recovery method to the proposed cubic structural element, all out-of-plane stress components can be recovered within a unified framework. However, the implementation requires transferring the stresses from the adjacent shell elements to the cohesive element, which substantially increases the complexity of the user element (UEL) subroutine implementation.
In a different approach, \citet{russo2020overcoming} employed the cohesive tractions as the out-of-plane stresses. However, this method was developed only for two-dimensional models and may produce excessive compressive stresses ahead of the cohesive zone.


Inspired by these studies, the present work directly employs the cohesive tractions obtained from three-dimensional cohesive elements to recover the out-of-plane stress components. 
One important parameter of the proposed method is the penalty stiffness of the cohesive element. \citet{turon2007engineering} proposed a penalty stiffness based on the laminate elastic modulus and thickness. However, the formulation is predicated on the assumption that the cohesive element has zero thickness. In practice, the shell-based cohesive element located between the two shell elements possesses a finite effective thickness, which equals one-half of the sum of the thicknesses of the adjacent shell elements~\cite{davila2007cohesive,davila2008effective}. Therefore, the penalty stiffness predicted by Turon's formulation is significantly higher than the actual stiffness of structural cohesive elements. The calculated stress distribution overestimates the true stress levels. This discrepancy motivates the development of a more physically representative penalty stiffness formulation.
\citet{polla2021delamination} proposed a method for evaluating the penalty stiffness of cohesive elements inserted between shell elements. However, the cohesive element thickness is defined as the sum of the thicknesses of the top and bottom shell elements, rather than one-half of the sum. \citet{bazilevs2018new} proposed an alternative penalty stiffness formulation based on the shell thickness. In this formulation, the cohesive element thickness is defined as one-half of the sum of the thicknesses of the adjacent shell elements. However, this formulation considers only the shear penalty stiffness and does not account for the dependence of the normal penalty stiffness on the shell thickness.
However, this formulation considers only the tangential penalty stiffness and does not account for the dependence of the normal penalty stiffness on the shell thickness. Both methods assume that the shell undergoes simple through-thickness shear deformation under transverse shear loading. This assumption is inconsistent with the Kirchhoff–Love shell theory, in which the shell normal remains straight and perpendicular to the mid-plane during deformation. Both studies focused on improving delamination simulations, whereas the recovery of a complete three-dimensional stress field was not considered. \citet{magisano2024large} proposed a hierarchical Kirchhoff–Love shell model with a warping model by introducing a soft layer between two stiff layers. With an appropriate warping model, the formulation accurately reproduces the response predicted by three-dimensional solid models. However, it is limited to the prediction of displacements, stability, in-plane stresses in the stiff layers, and transverse shear deformation in the soft interlayer. Predictions of out-of-plane stresses were not presented.

In this paper, a new penalty stiffness formulation for recovering out-of-plane stresses in layer-wise shell models of composite laminates is presented. The proposed formulation expresses the penalty stiffness as a function of the mechanical properties and the thickness of the resin-rich regions between adjacent plies. The objective of this work is to extend the previously developed structural element model for composite laminates while satisfying the following requirements:
\begin{enumerate}
    \item It retains the key advantage of the previous formulation by eliminating the cohesive-zone mesh-density limitation in delamination analysis;
    \item It can accurately and efficiently recover the three-dimensional stress field in an arbitrary multi-ply composite laminate. 
\end{enumerate}

The rest of the paper is organized as follows: Section~\ref{sec:method} presents the proposed method in detail. Section~\ref{sec:Numerical examples} demonstrates the performance of these elements on a series of uniaxial extension benchmarks for both cross-ply and angle-ply laminates. In the end, Section~\ref{sec:summary} draws the conclusions of this work and discusses some potential future work.

\section{Method}\label{sec:method}
\begin{figure}[h!]
	\centering
	\includegraphics[width=1.0\linewidth]{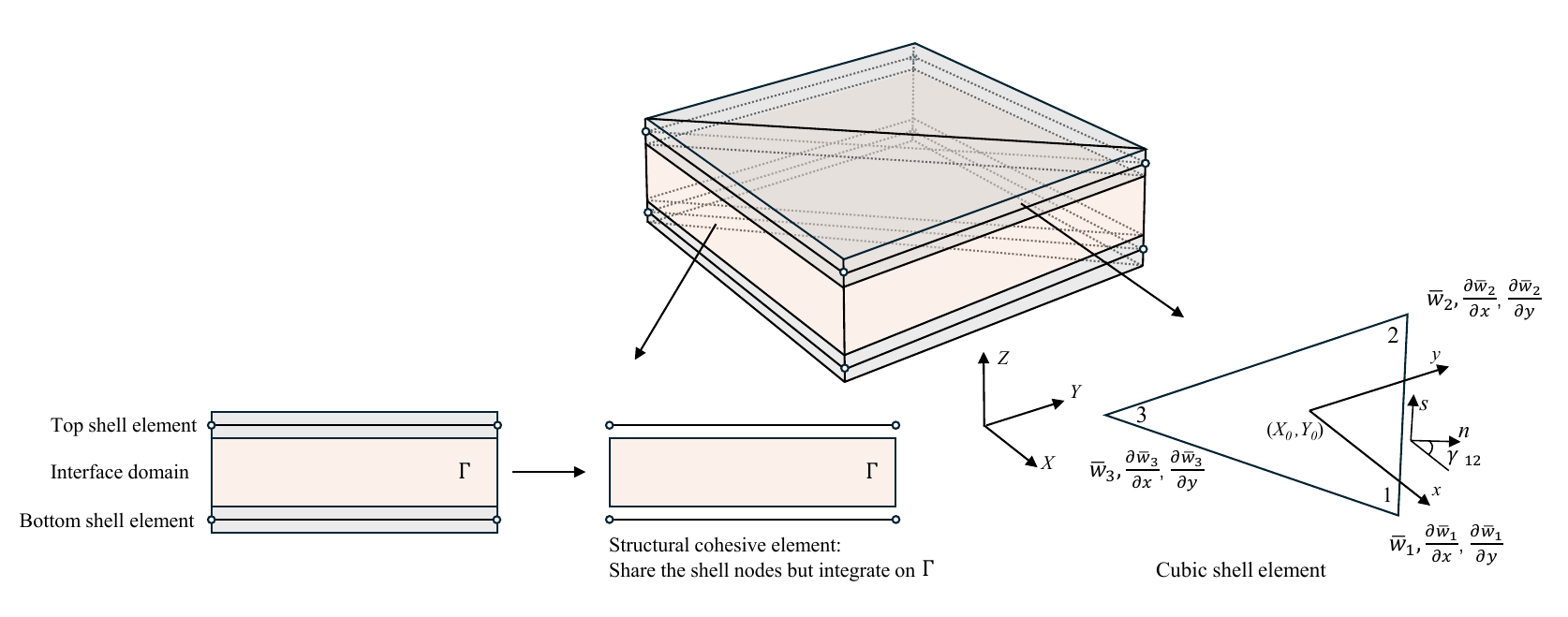}
	\caption{Triangular cubic structural element}
	\label{fig:Structural element}
\end{figure}

The proposed formulation is based on the Kirchhoff–Love hypothesis and employs a layer-wise representation of composite laminates. Each ply is modeled using a cubic shell element, while adjacent plies are connected by structural cohesive elements, as shown in Figure~\ref{fig:Structural element}. The detailed derivation of the cubic shell element and the compatible structural cohesive element formulations is presented in \citet{ai2025structural}.

\subsection{3D stress calculation in cohesive elements}
\label{subsec-3D stress calculation}
Most 3D composite failure criteria, such as those proposed by Hashin~\cite{hashi1980failure}, Puck~\cite{puck2004failure}, and Pinho~\cite{pinho2006physically,chen2013numerical}, require the complete stress state within each ply. However, cubic shell elements based on the Kirchhoff--Love hypothesis provide only the in-plane stress components. The out-of-plane stresses can be recovered from the tractions transmitted by the structural cohesive elements.

\begin{figure}[h!]
	\centering
    \includegraphics[width=0.95\linewidth]{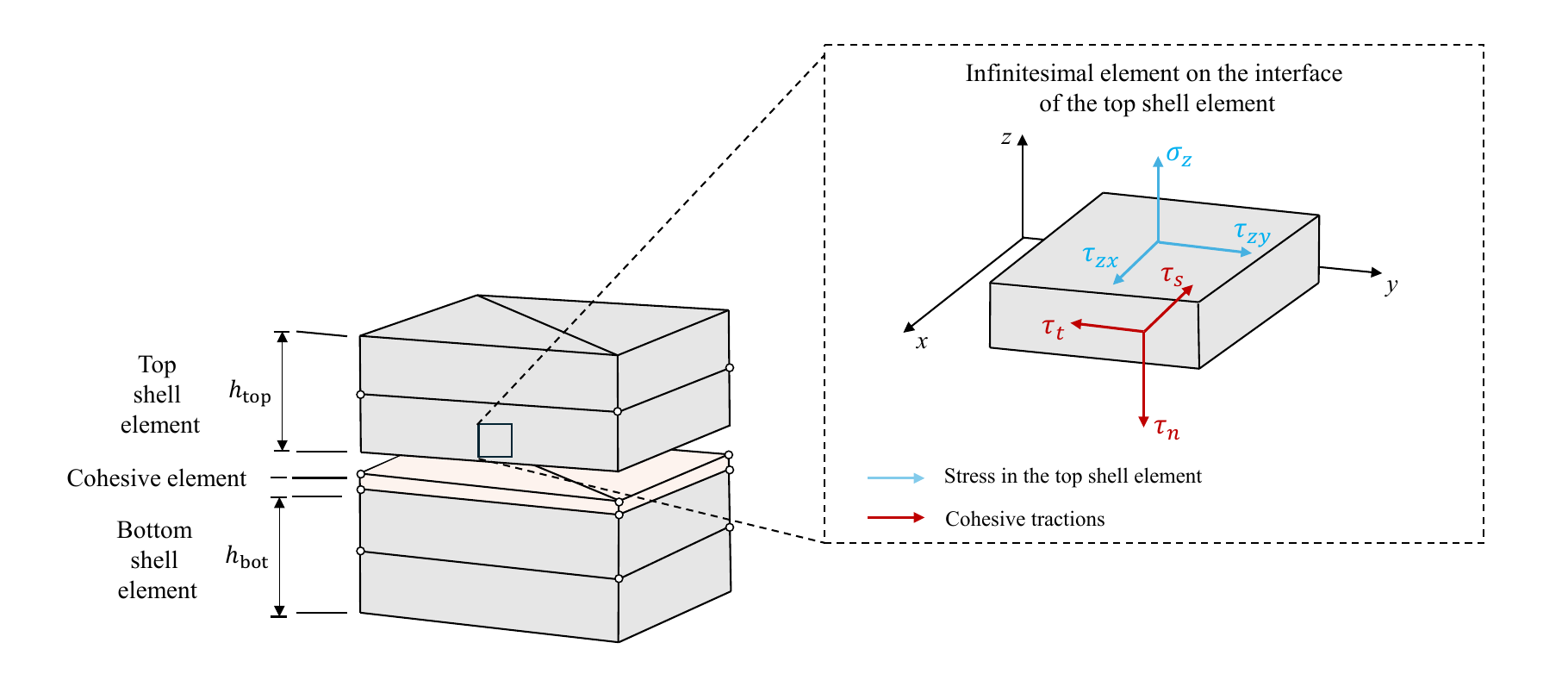}
	\caption{Coordinate system and DoFs for the triangular cubic plate element}
	\label{fig:3d stress}
\end{figure}

Based on the shell thickness and kinematic equations, it is possible to determine the displacements at the interfaces of two adjacent shell elements. Consider an infinitesimal element located adjacent to the interface of the upper shell element, as illustrated in Figure~\ref{fig:3d stress}. Three stress components act on the surface in contact with the cohesive element, namely the out-of-plane normal stress (denoted as $\sigma_z$) and the two transverse shear stresses (denoted as $\tau_{xz}$ and $\tau_{zy}$).By equilibrium, these stresses are equal in magnitude and opposite in direction to the corresponding cohesive tractions($\tau_n$, $\tau_s$, and $\tau_t$), respectively.
\begin{align}
    &\sigma_z = -\tau_n \\[0.5em]
    &\tau_{zx} = -\tau_s \\[0.5em]
    &\tau_{zy} = -\tau_t  
\end{align}

Therefore, the normal and shear tractions transmitted by the cohesive elements provide the out-of-plane components of the three-dimensional stress tensor.

\subsection{Interpolation method of stress distribution}
\label{subsec-Interpolation along thickness}

\begin{figure}[h!]
	\centering
	\includegraphics[width=0.85\linewidth]{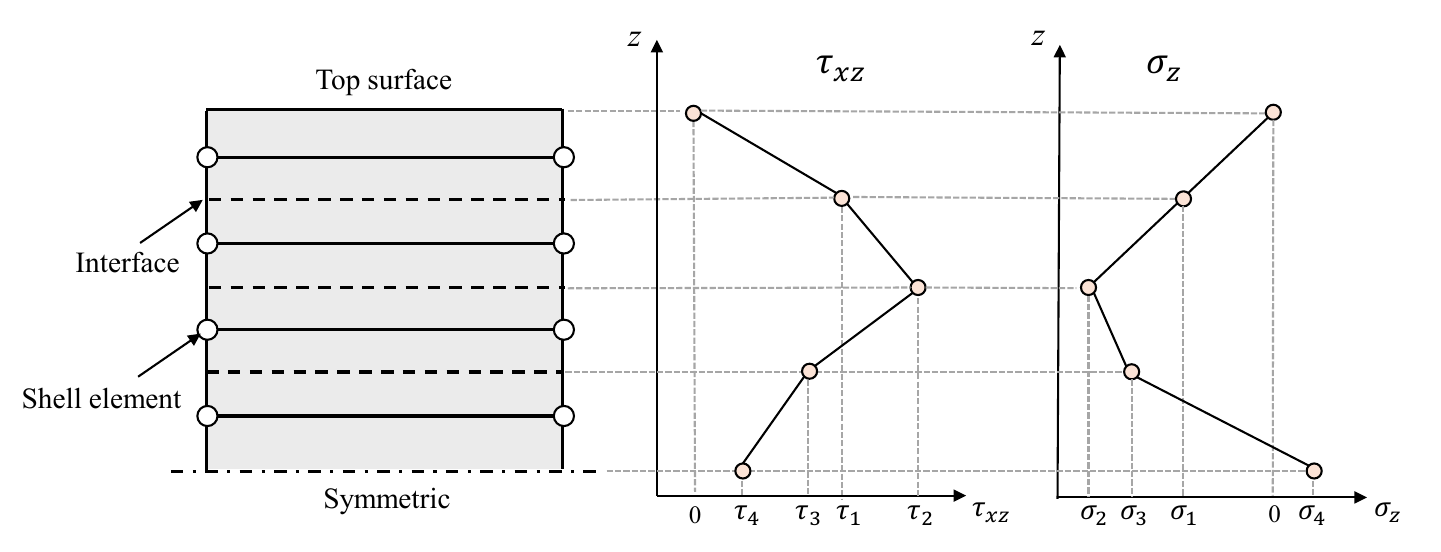}
	\caption{Interpolation method of the structural elements along thickness}
	\label{fig:3d Interpolation-thickness}
\end{figure}

The cohesive tractions described in Section~\ref{subsec-3D stress calculation} are evaluated at the integration points of the cohesive elements. In this work, each cohesive element contains 13 integration points. Since the traction values are not directly available at the nodes, they are obtained by substituting the node coordinates into the shape function of the element. As each node is shared by multiple adjacent elements, the nodal traction is calculated by averaging the extrapolated values from all neighboring elements associated with this node. Following this procedure, the interfacial stresses at every ply interface can be determined. The through-thickness distributions of the out-of-plane normal and shear stresses are then derived by interpolation. For simplicity, a linear interpolation is adopted between adjacent cohesive interfaces, as shown in Figure~\ref{fig:3d Interpolation-thickness}. The implementation and validation of this interpolation procedure are presented in Section~\ref{sec:Numerical examples}.

\subsection{Penalty stiffness of cohesive elements}
\label{subsec-Penalty stiffness}
The cohesive elements are characterized based on a proper traction separation law, which describes a constitutive relation linking the opening displacements and stresses under the three fracture modes. In this work, a bilinear traction-separation law is adopted. The law is defined by three parameters: the interface strength, $\tau$, the fracture toughness, $G_c$, and the penalty stiffness, $K$.
The interface strength and fracture toughness are used for damage initiation and evolution, whereas the penalty stiffness determines the elastic response before damage initiation. Since the present study focuses on the recovery of the three-dimensional stress field, all cohesive elements are assumed to remain undamaged. Accordingly, the displacement, $\delta$, is required to remain below the damage initiation displacement, $\Delta_0$.

As reviewed in Section~\ref{sec:Introduction}, \citet{turon2007engineering} proposed an analytical formulation for determining the cohesive penalty stiffness. In their formulation, zero-thickness cohesive elements are inserted between adjacent three-dimensional solid elements. The penalty stiffness is derived as a function of the material parameters and a coefficient, $\alpha$, which is defined from the experimental results:
\begin{align}
    K_n &= \frac{\alpha E_3}{h} \\[0.5em]
    K_s &= K_t = \frac{\alpha G_{13}}{h} 
\end{align}
where $h$ is the thickness of the laminate and coefficient $\alpha$ typically assumes a value of 50~\cite{turon2007engineering}. 

\begin{figure}[htbp]
	\centering
	\includegraphics[width=0.9\linewidth]{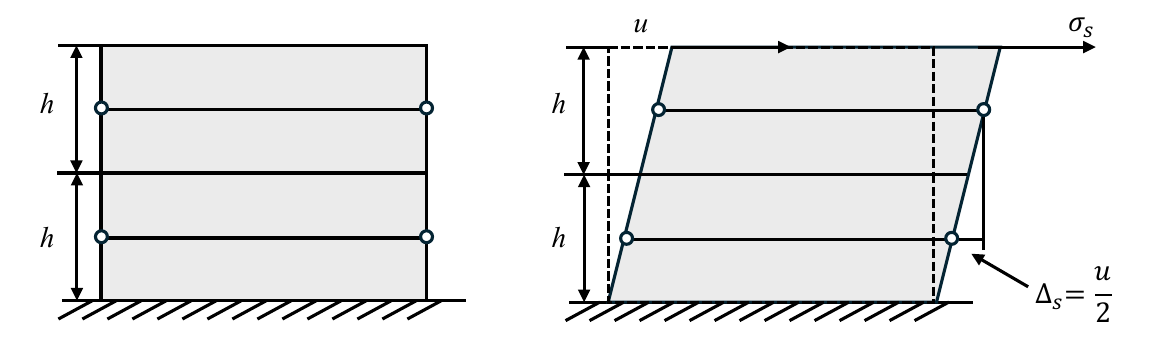}
	\caption{State of simple shear in the cohesive model}
	\label{fig:simple shear-Ba}
\end{figure}

Subsequently, \citet{bazilevs2018new} proposed a modified formulation for the penalty stiffness based on the Reissner-Mindlin theory. The formulation is derived by reproducing the deformation of solid elements under simple shear loading, as shown in Figure~\ref{fig:simple shear-Ba}:
\begin{align}\label{eq:ks=kt-Bk}
    K_s = K_t = \frac{G_{13}}{h_{\text{top}}/2 +h_{\text{bot}}/2}
\end{align}
where $h_{\textrm{top}}$ and $h_{\textrm{bot}}$ correspond to the thicknesses of the top and bottom laminates, respectively. As discussed in Section~\ref{sec:Introduction}, no corresponding formulation was proposed for the normal penalty stiffness. Equation~\ref{eq:ks=kt-Bk} is derived under the assumption of simple shear deformation of the solid element and is therefore inconsistent with the kinematic assumptions of the Kirchhoff–Love shell theory. The formulations proposed in this work for the normal and shear penalty stiffnesses are presented in the following sections.

\subsubsection{Normal penalty stiffness, $K_n$}\label{subsubsec:normal}
\begin{figure}[htbp]
	\centering
	\includegraphics[width=0.9\linewidth]{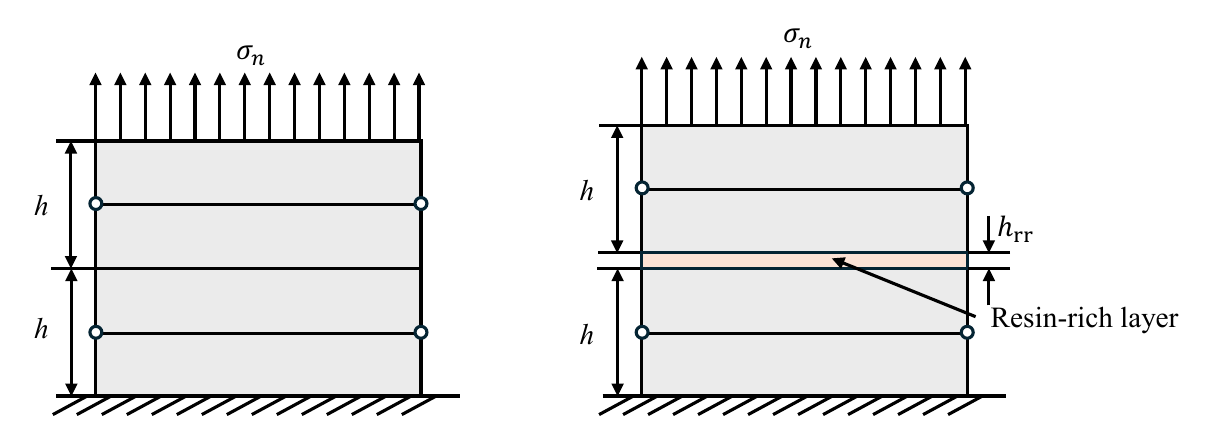}
	\caption{State of tensile with resin-rich layer}
	\label{fig:K-modeI}
\end{figure}

Consider a structural element subjected to a uniform normal tensile traction acting on the top surface of the shell element, as illustrated in Figure~\ref{fig:K-modeI}. Assuming that the deformation is concentrated in the resin-rich layer, the out-of-plane normal stress can be expressed as:
\begin{align}\label{eq:normal-stress-strain}
    \sigma_n &= E_3\varepsilon_z  \nonumber\\[0.5em]
     &= \frac{E_3w}{2h_{\textrm{rr}}}
\end{align}
where $w$ is the total normal displacement of shell elements, $E_3$ and $h_\textrm{rr}$ are the out-of-plane elastic modulus and thickness of the resin-rich layer, respectively.

The cohesive normal jump displacement, $\Delta_n$, is defined as the displacement difference between the nodes of the adjacent shell elements rather than the displacement at the shell surfaces. Since the displacement $w$ represents the displacement between the shell surfaces, the normal jump displacement can be expressed as:
\begin{align}\label{eq:jump-disp-normal}
    \Delta_n = \frac{w}{2}
\end{align}

Combining Equations~\ref{eq:normal-stress-strain} and \ref{eq:jump-disp-normal}, the normal penalty stiffness is obtained as:
\begin{align}\label{eq:Kn-hrr}
    K_n &= \frac{\sigma_n}{\Delta_n} \nonumber \\[0.5em]
    &= \frac{E_3}{h_\textrm{rr}}
\end{align}

\subsubsection{Shear penalty stiffness, $K_s$ and $K_t$}\label{subsubsec:shear}
\begin{figure}[h]
	\centering
	\includegraphics[width=0.9\linewidth]{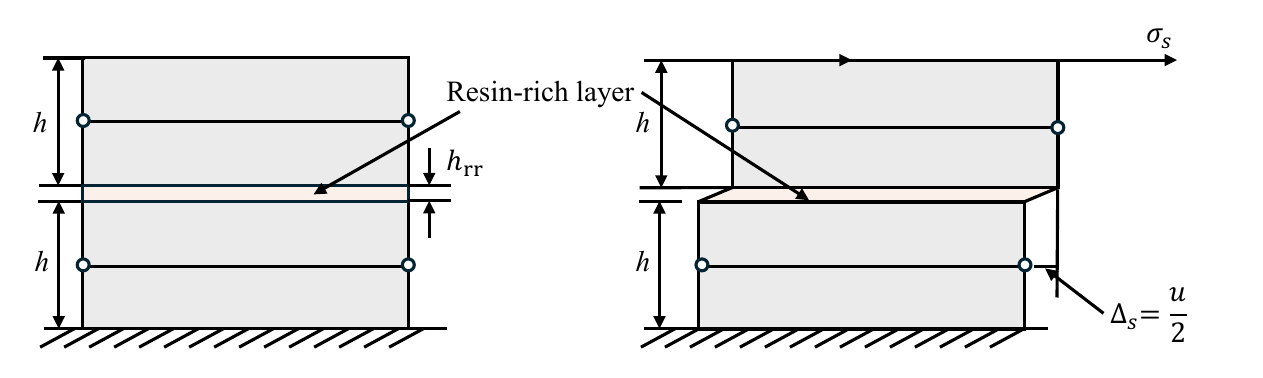}
	\caption{State of simple shear with resin-rich layer}
	\label{fig:K-modeII}
\end{figure}

To determine the shear penalty stiffnesses, $K_s$ and $K_t$, a different kinematic assumption from those adopted in previous studies is employed. Consistent with the Kirchhoff-Love hypothesis, the shell normals remain perpendicular to the mid-plane throughout the deformation.  The deformation produced by transverse shear loading is assumed to occur only within the resin-rich region, as illustrated in Figure~\ref{fig:K-modeII}. Assuming the resin-rich layer material is isotropic, the transverse shear strain is uniform and can be expressed as:
\begin{align}
    \gamma = \frac{u}{2h_{\textrm{rr}}}
\end{align}
The corresponding transverse shear stress is given by:
\begin{align}
    \sigma_{s} = G\gamma = \frac{Gu}{2h_{\textrm{rr}}}
\end{align}
where $G$ is the shear modulus of the resin-rich layer. Similar to the tensile case, the cohesive shear jump displacement $\Delta_s$ is:
\begin{align}
    \Delta_s = \frac{u}{2}
\end{align}

Combining the above relations, the shear penalty stiffness is obtained as
\begin{align}\label{eq:Ks-hrr}
    K_s &= \frac{\sigma_s}{\Delta_s} \nonumber \\[0.5em]
    &= \frac{G}{h_{\textrm{rr}}}
\end{align}

\section{Numerical examples}\label{sec:Numerical examples}
As discussed in Section~\ref{sec:Introduction}, the high stress gradients near the free edge are commonly attributed to stress singularities. This section aims to verify the proposed formulation by recovering the three-dimensional stress field and evaluating its ability to predict free-edge stresses. To achieve these objectives, a uniform axial extension (UAE) model is adopted for verification. For cross-ply and angle-ply laminates, the predicted stress distributions are compared with analytical solutions and finite element results obtained using C3D8 solid elements in Abaqus. For general laminate lay-ups, the comparisons are performed only against the C3D8 solid-element model. The detailed results are presented in the following sections.
\subsection{Uniform axial extension model}
\label{subsec:UAE model}
\begin{figure}[htbp]
      \centering
	   \begin{subfigure}{0.45\linewidth}
		\includegraphics[width=\linewidth]{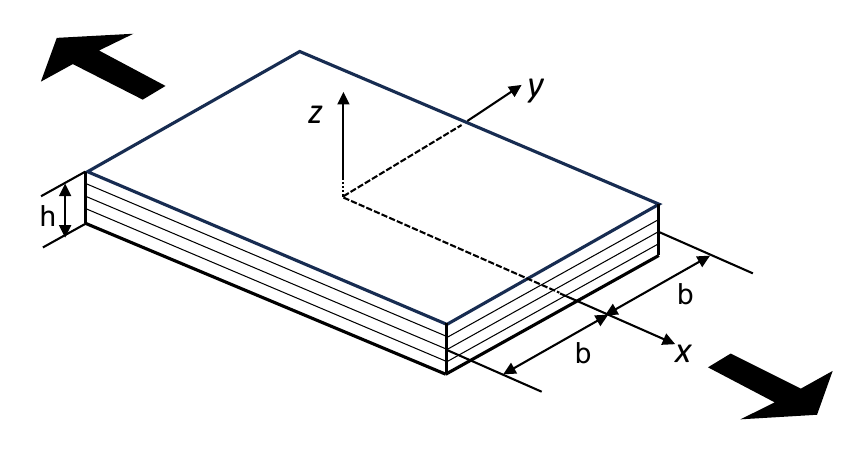}
		\caption{4-ply model}
		\label{fig:UAE_4L}
	   \end{subfigure}
	   \begin{subfigure}{0.45\linewidth}
		\includegraphics[width=\linewidth]{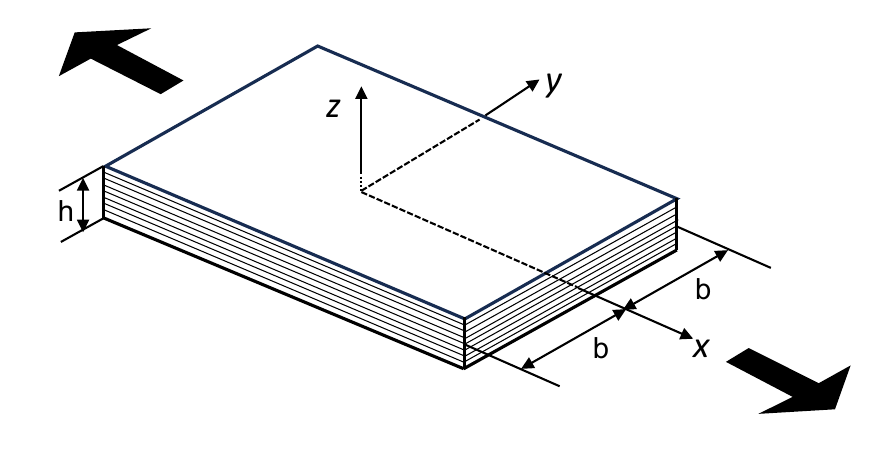}
		\caption{8-ply model}
		\label{fig:UAE_8L}
	    \end{subfigure}
	\caption{{Geometry of the uniform axial extension laminate model}}
	\label{fig:UAE_4L_8L}
\end{figure}

The geometric configuration of the composite laminate is shown in Figure~\ref{fig:UAE_4L_8L}. The laminate has a total thickness of $h$, a width of $2b$, and a length of $L$. It consists of multiple plies, each with a thickness of $h_0$. For the four-ply laminates considered in Sections~\ref{subsec-cross-ply} and~\ref{subsec-angle-ply-theta}, the half-width is taken as $b=4h_0$. For the eight-ply laminates, $b=16h_0$. Each ply is modeled using a single layer of cubic shell elements, while adjacent plies are connected by cubic structural cohesive elements. The origin of the global coordinate system is located at the center of the laminate. The X-axis is along the length of the model, the Y-axis is along the width of the model, and the Z-axis is in the through-thickness direction of the model. A uniform displacement, $u$, is applied in the longitudinal direction. Rigid-body motion is prevented by imposing pin constraints at the center of the laminate.

\subsection{Postprocessing technique for stress distribution along the width of laminates}
\begin{figure}[h]
	\centering
	\includegraphics[width=1.0\linewidth]{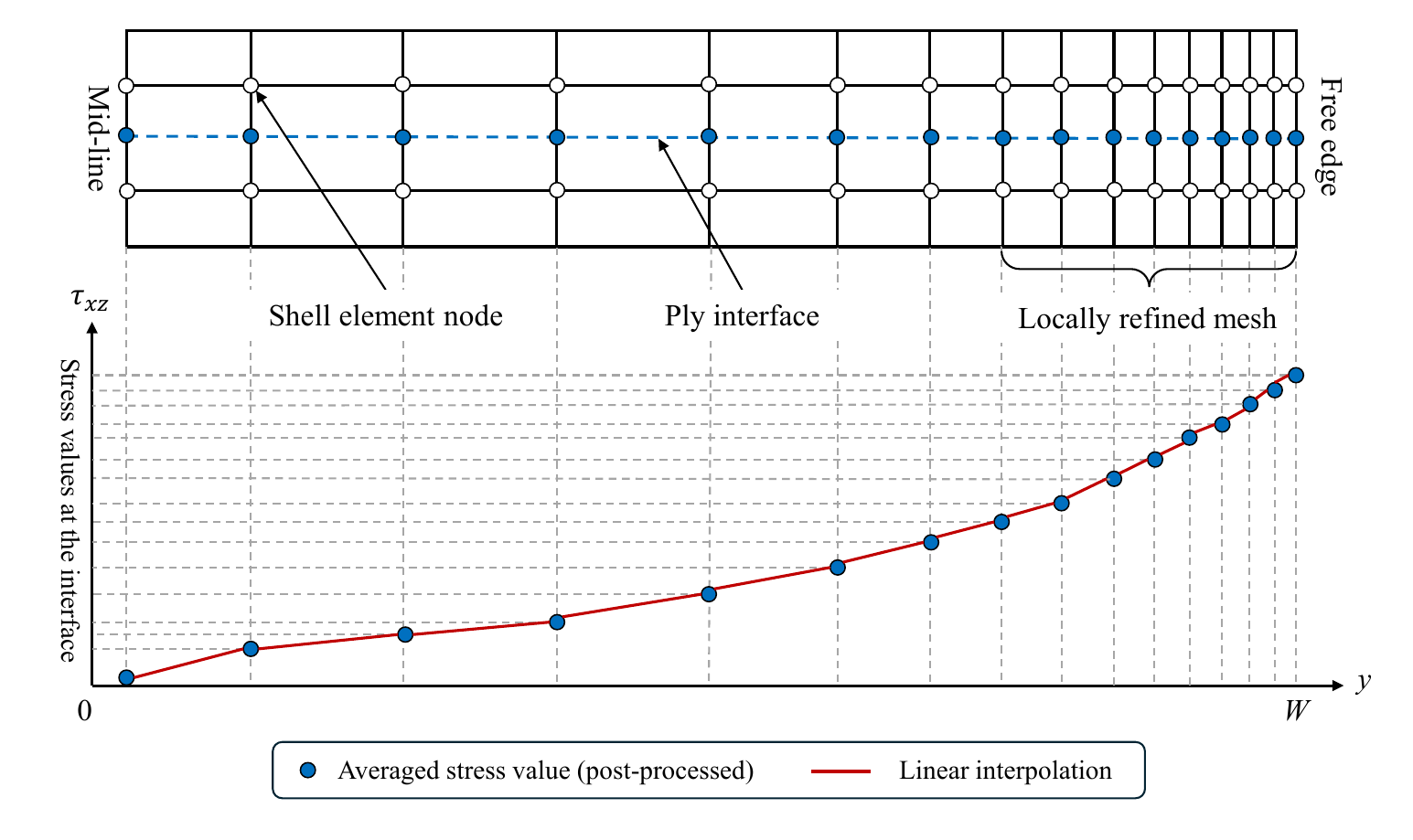}
	\caption{Interpolation method of the structural elements along the width}
	\label{fig:3d Interpolation-width}
\end{figure}

To obtain the stress distribution along the laminate width at a given ply interface, the mesh is locally refined near the free edge. The nodal stresses are then evaluated using the averaging procedure described in Section~\ref{subsec-Interpolation along thickness}. The resulting nodal values are plotted and connected by straight line segments to reconstruct the stress distribution. The post-processing procedure for the shear stress ($\tau_{xz}$) distribution at the mid-plane is illustrated in Figure~\ref{fig:3d Interpolation-width}.

\subsection{Results for cross-ply laminates $[0^{\circ}/90^{\circ}]_{s}$}
\label{subsec-cross-ply}
Cross-ply $[0^{\circ}/90^{\circ}]_s$ laminate has been widely used as a benchmark for free-edge stress analysis~\cite{wang1977some,wang1978interlaminar}. In the present study, it is adopted to evaluate the out-of-plane normal stress, $\sigma_z$, and transverse shear stress, $\tau_{yz}$, at the ($0^{\circ}/90^{\circ}$) interface and the laminate mid-plane. The geometry of the 4-ply laminate model is illustrated in Figure~\ref{fig:UAE_4L_8L}, and the corresponding geometric parameters are presented in Table~\ref{tab:4-layer-geo}. The material properties of the cross-ply laminate are summarized in Table~\ref{tab:Crossply_Material}.
\begin{table}[htbp]
  \centering
  \caption{Geometric parameters for the four-ply laminate model}
    \begin{tabular}{llll}
    \toprule
    Parameter &  $h_0$ & $h$  & $b$ (width)  \\[0.3em]
 \midrule
    value (mm) & 0.125 & 0.5 & 1  \\[0.3em]
    \bottomrule
    \end{tabular}%
  \label{tab:4-layer-geo}%
\end{table}

The elastic modulus and Poisson's ratio of the resin-rich layer are taken from \citet{burhan2024three}, while the thickness of the resin-rich layer is taken from \citet{grande1991effects}. The corresponding penalty stiffnesses are calculated using Equations.~\ref{eq:Kn-hrr} and~\ref{eq:Ks-hrr}. The resulting values are listed in Table~\ref{tab:Kvalue-RRL}. It should be noted that the exact properties of the resin-rich layer are unavailable for the laminates considered. The values reported in Table~\ref{tab:Kvalue-RRL} are approximate and are estimated from similar material. The influence of these parameters is investigated through a sensitivity analysis in the following section.
\begin{table}[h!]
  \centering
  \caption{Material properties of cross-ply laminate \cite{pagano1978stress}}
    \begin{tabular}{lll}
    \toprule
 \multicolumn{3}{l}{\textbf{Material properties of $[0^{\circ}/90^{\circ}]_s$ laminate} }   \\
 \midrule
    $E_{11}$= 137900 MPa & $E_{22}$= 14500 MPa & $E_{33}$ = 14500 MPa \\[0.3em]
    $\nu_{12}$ = 0.21 & $\nu_{13}$ = 0.21 & $\nu_{23}$ = 0.21  \\[0.3em]
    $G_{12}$ = 5900 MPa & $G_{13}$ = 5900 MPa & $G_{23}$ = 5900 MPa \\[0.5em]
    \bottomrule
    \end{tabular}%
  \label{tab:Crossply_Material}%
\end{table}%

\begin{table}[htbp]
  \centering
  \caption{Penalty stiffness obtained by resin-rich layer}
    \begin{tabular}{cccccccc}
    \toprule
    \thead{$E$\\ $\mathrm{(N/mm^2)}$} & $\nu$ & \thead{RRL thickness \\$(10^{-3}$inch)} &\thead{RRL thickness \\(mm)}&   \thead{$\K_n$ \\ $\mathrm{(N/mm^3)}$} & \thead{$\K_t$ \\ $\mathrm{(N/mm^3)}$} &  \thead{$\K_s$ \\ $\mathrm{(N/mm^3)}$}  \\[0.3em]
 \midrule
    3000 & 0.3 & 0.9 & 0.02286 & 131233 & 50474 & 50474 \\[0.3em]
    \bottomrule
    \end{tabular}%
  \label{tab:Kvalue-RRL}%
\end{table}

\subsubsection{Numerical results obtained by the proposed penalty stiffness}

\begin{figure}[h!]
	\centering
	\includegraphics[scale=1]{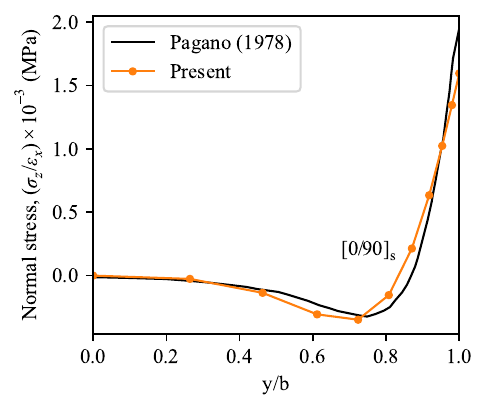}
	\caption{Interlaminar normal stress $\sigma_z$ at the mid-plane}
	\label{fig:0/90-sigmaZ-z=0-0.9inch}
\end{figure}

Figure~\ref{fig:0/90-sigmaZ-z=0-0.9inch} compares the interlaminar normal stress, $\sigma_z$, at the laminate mid-plane. The analytical solution is compared with the proposed structural-element model for a resin-rich layer thickness of $0.9\times10^{-3}$ inch. The black solid line represents the analytical solution reported by \citet{pagano1978stress}, while the orange solid line corresponds to the results obtained using the proposed structural elements. It can be found that the numerical results agree well with the analytical solution for $y/b<0.5$. As $y/b$ exceeds 0.5, a small deviation is observed. Specifically, the maximum stress at the free edge predicted by the proposed method is slightly lower than the analytical solution.

\begin{figure}[h!]
	\centering
	\includegraphics[scale=1]{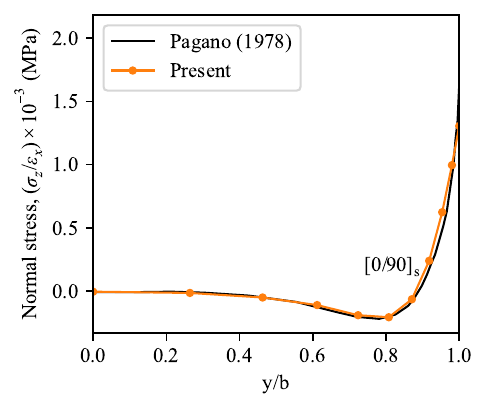}
	\caption{Interlaminar normal stress $\sigma_z$ at the interface between $0^{\circ}$ and $90^{\circ}$}
	\label{fig:0/90-sigmaZ-z=h0-0.9inch}
\end{figure}

Figure~\ref{fig:0/90-sigmaZ-z=h0-0.9inch} compares the numerical and analytical distributions of the interlaminar normal stress, $\sigma_z$, at the $(0^{\circ}/90^{\circ})$ interface. The agreement is even better than that obtained at the laminate mid-plane. For $y/b<0.9$, the numerical results are identical to the analytical solution. Unlike the mid-plane results, the maximum stress at the free edge is also accurately predicted, remaining very close to the analytical solution.

\begin{figure}[h!]
	\centering
	\includegraphics[scale=1]{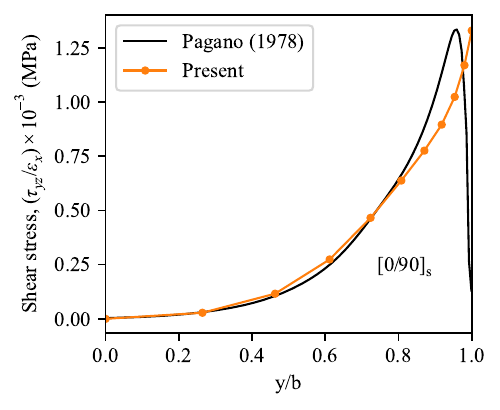}
	\caption{Interlaminar shear stress $\tau_{yz}$ at the interface between $0^{\circ}$ and $90^{\circ}$}
	\label{fig:0/90-sigmaZ-z=h0-0.9-inch-tau}
\end{figure}

Finally, Figure~\ref{fig:0/90-sigmaZ-z=h0-0.9-inch-tau} compares the interlaminar shear stress, $\tau_{yz}$, at the $(0^{\circ}/90^{\circ})$ interface. Overall, the numerical and analytical solutions are in good agreement. An obvious discrepancy is observed only at the free edge, where the numerical stress does not reduce to the zero value predicted by the analytical solution. This difference is attributed to the nodal interpolation procedure employed in the post-processing stage.

The above results demonstrate that, when the resin-rich layer properties listed in Table~\ref{tab:Kvalue-RRL} are adopted, the predicted stress distributions are in good agreement with the analytical solutions. In practical applications, the material properties and thickness of the resin-rich layer are generally uncertain. As shown in Equations~\ref{eq:Kn-hrr} and~\ref{eq:Ks-hrr}, the penalty stiffness is governed primarily by these two parameters. A sensitivity analysis is conducted to evaluate their influence on the predicted stress distributions.

\subsubsection{Sensitivity analysis: Resin-rich layer thickness}
\citet{grande1991effects} measured the average thickness of the resin-rich layer through quantitative analysis of polished cross-sectional micrographs. Three representative resin-rich layer thicknesses, $0.6\times10^{-3}$, $0.7\times10^{-3}$, and $0.9\times10^{-3}$ inch, were reported. The corresponding penalty stiffnesses were calculated for each thickness using the proposed formulation, and the resulting values are summarized in Table~\ref{tab:Kvalue-Hrr-different-Thickness}.
\begin{table}[htbp]
  \centering
  \caption{Normal and shear penalty stiffness values for different resin-rich layer thicknesses}
    \begin{tabular}{cccccc}
    \toprule
     Model name & \thead{RRL thickness \\($10^{-3}$inch)} &\thead{RRL thickness \\(mm)}&  \thead{$\K_n$ \\ $\mathrm{(N/mm^3)}$} & \thead{$\K_t$ \\ $\mathrm{(N/mm^3)}$} &  \thead{$\K_s$ \\ $\mathrm{(N/mm^3)}$}  \\[0.3em]
 \midrule
    T-0.9 & 0.9 & 0.02286 & 131233 & 50474 & 50474 \\[0.3em]
\midrule  
    T-0.7 & 0.7 & 0.01778 & 168728 & 64895 & 64895 \\[0.3em]
\midrule  
    T-0.6 & 0.6 & 0.01524 & 196850 & 75711 & 75711 \\[0.3em]    
    \bottomrule
    \end{tabular}%
  \label{tab:Kvalue-Hrr-different-Thickness}%
\end{table}

\begin{figure}[h!]
       \centering
	   \begin{subfigure}{0.48\linewidth}
		\includegraphics[scale=1]{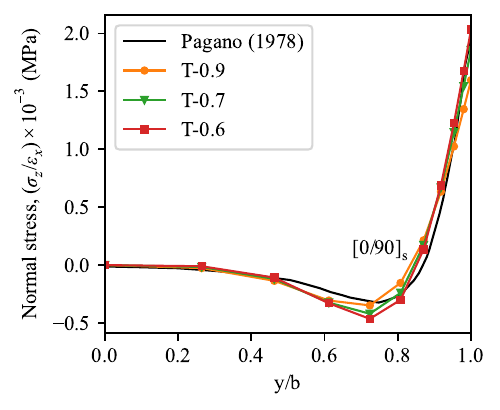}
		\caption{$\sigma_z$ at the mid-plane}
		\label{fig:Cross-ply-Hrr-mid}
	   \end{subfigure}
    ~
	     \begin{subfigure}{0.48\linewidth}
		 \includegraphics[scale=1]{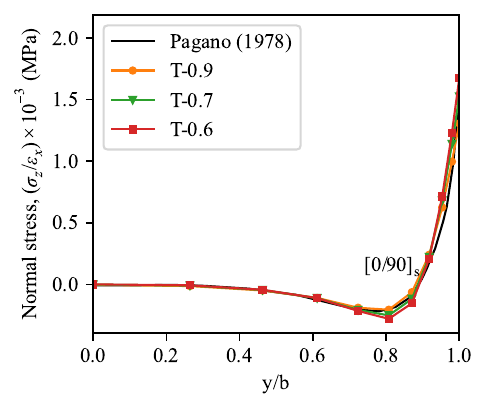}
		 \caption{$\sigma_z$ at the interface between $0^{\circ}$ and $90^{\circ}$}
		 \label{fig:Cross-ply-Hrr-H-sigmaZ}
	      \end{subfigure}
      \vfill  
      \begin{subfigure}{0.48\linewidth}
		 \includegraphics[scale=1]{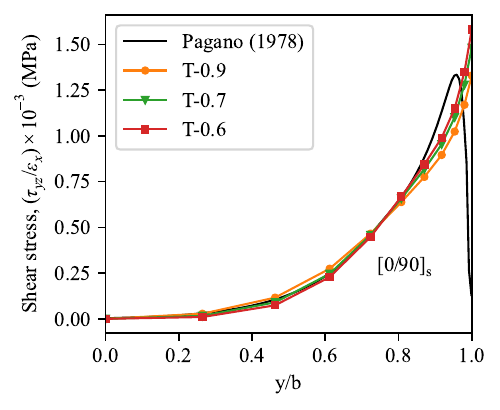}
		 \caption{$\tau_{yz}$ at the interface between $0^{\circ}$ and $90^{\circ}$}
		 \label{fig:Cross-ply-Hrr-H-tauYZ}
	      \end{subfigure}
	\caption{Distributions of interlaminar normal and shear stresses with different resin-rich layer thicknesses}
	\label{fig:Cross-ply-Hrr}
\end{figure}

The T-0.6 and T-0.7 models have the same geometry and boundary conditions as the T-0.9 model. To evaluate the influence of the resin-rich layer thickness, the predicted stress distributions are compared with the analytical solution and the results obtained from the T-0.9 model. The comparisons include the interlaminar normal stress, $\sigma_z$, at the laminate mid-plane and the $(0^{\circ}/90^{\circ})$ interface, as well as the interlaminar shear stress, $\tau_{yz}$, at the $(0^{\circ}/90^{\circ})$ interface. The corresponding results are presented in Figure~\ref{fig:Cross-ply-Hrr}.

Only small differences are observed among the three models, and all predictions remain in close agreement with the analytical solution. This demonstrates that the proposed penalty stiffness formulation is insensitive to the variation in resin-rich layer (RRL) thickness within the range reported by \citet{grande1991effects}. In practical applications, once the material properties and thickness of the RRL are measured, the corresponding penalty stiffnesses can be determined directly from Equations~\ref{eq:Kn-hrr} and~\ref{eq:Ks-hrr}. Based on these results, an RRL thickness of $0.9\times10^{-3}$ inch is adopted in the subsequent validation studies.

\subsubsection{Sensitivity analysis: Elastic modulus and Poisson's ratio}
In the study by \citet{burhan2024three}, the resin-rich layer was assumed to have an elastic modulus of $E=3\,\text{GPa}$ and a Poisson's ratio of $\nu=0.3$. In practice applications, the mechanical properties of the resin-rich layer vary with the epoxy resins employed~\cite{krueger2006analysis,gupta2015potential,li2019compressive}. \citet{gupta2015potential} reported an elastic modulus ranging from $2.7$ to $4.1\,\text{GPa}$. \citet{krueger2006analysis} reported $E=4.67\,\text{GPa}$ and $\nu=0.37$ for the 8552 epoxy resin in the IM7/8552 composite. Based on these data, this study adopts ranges of $ 3{-}5 \, \text{GPa} $ for the resin-rich layer elastic modulus and $0.2{-}0.4$ for Poisson’s ratio in the sensitivity analysis. The corresponding penalty stiffnesses are summarized in Tables~\ref{tab:Kvalue-Hrr-different-E} and~\ref{tab:Kvalue-Hrr-different-Nu}, respectively.

\begin{table}[htbp]
  \centering
  \caption{Normal and shear penalty stiffness values for different resin-rich layer elastic moduli}
    \begin{tabular}{cccccc}
    \toprule
     Model name & \thead{Elastic modulus \\$\mathrm{(N/mm^2)}$} &\thead{RRL thickness \\$(10^{-3}$inch)}&  \thead{$\K_n$ \\ $\mathrm{(N/mm^3)}$} & \thead{$\K_t$ \\ $\mathrm{(N/mm^3)}$} &  \thead{$\K_s$ \\ $\mathrm{(N/mm^3)}$}  \\[0.3em]
 \midrule
    E-3000 & 3000 & 0.9 & 131233 & 50474 & 50474 \\[0.3em]
\midrule  
    E-4000 & 4000 & 0.9 & 174978 & 67299 & 67299 \\[0.3em]
\midrule  
    E-5000 & 5000 & 0.9 & 218722 & 84124 & 84124 \\[0.3em]    
    \bottomrule
    \end{tabular}%
  \label{tab:Kvalue-Hrr-different-E}%
\end{table}

\begin{table}[htbp]
  \centering
  \caption{Normal and shear penalty stiffness values for different resin-rich layer Poisson's ratios}
    \begin{tabular}{cccccc}
    \toprule
     Model name & Poisson's ratio &\thead{RRL thickness \\$(10^{-3}$inch)}&  \thead{$\K_n$ \\ $\mathrm{(N/mm^3)}$} & \thead{$\K_t$ \\ $\mathrm{(N/mm^3)}$} &  \thead{$\K_s$ \\ $\mathrm{(N/mm^3)}$}  \\[0.3em]
 \midrule
    Nu-0.3 & 0.3 & 0.9 & 131233 & 50474 & 50474 \\[0.3em]
\midrule  
    Nu-0.2 & 0.2 & 0.9 & 131233 & 54680 & 54680 \\[0.3em]
\midrule  
    Nu-0.4 & 0.4 & 0.9 & 131233 & 46869 & 46869 \\[0.3em]    
\bottomrule
    \end{tabular}%
  \label{tab:Kvalue-Hrr-different-Nu}%
\end{table}

\begin{figure}[h!]
       \centering
	   \begin{subfigure}{0.48\linewidth}
		\includegraphics[scale=1]{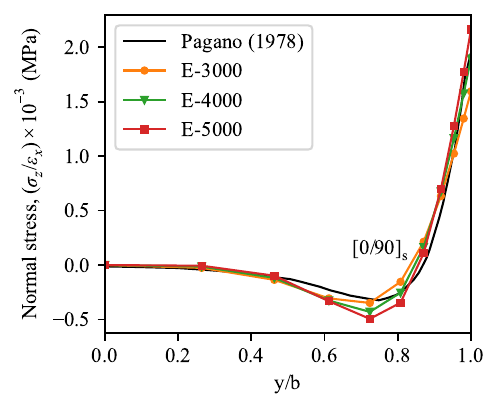}
		\caption{$\sigma_z$ at the central plane}
		\label{fig:Cross-ply-E-mid}
	   \end{subfigure}
    ~
	     \begin{subfigure}{0.48\linewidth}
		 \includegraphics[scale=1]{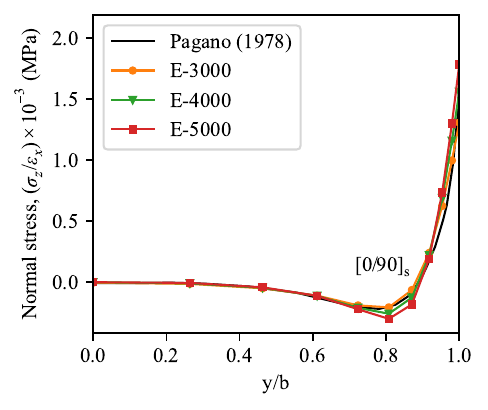}
		 \caption{$\sigma_z$ at the interface between $0^{\circ}$ and $90^{\circ}$}
		 \label{fig:Cross-ply-E-H-sigmaZ}
	      \end{subfigure}
      \vfill  
      \begin{subfigure}{0.48\linewidth}
		 \includegraphics[scale=1]{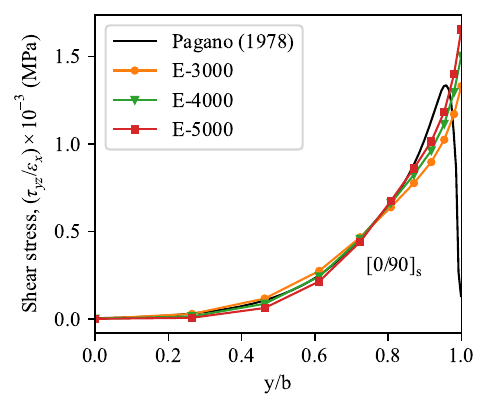}
		 \caption{$\tau_{yz}$ at the interface between $0^{\circ}$ and $90^{\circ}$}
		 \label{fig:Cross-ply-E-H-tauYZ}
	      \end{subfigure}
	\caption{Distributions of interlaminar normal and shear stresses with different elastic moduli}
	\label{fig:Cross-ply-E}
\end{figure}

Figure~\ref{fig:Cross-ply-E} presents the interlaminar normal stress ($\sigma_z$) and the interlaminar shear stress ($\tau_{yz}$) for different elastic moduli of the resin-rich layer. The stress distributions are evaluated at the laminate mid-plane and the $(0^{\circ}/90^{\circ})$ interface over the range $0\leq y/b\leq1$. As shown in the figure, only small differences are observed among the different elastic modulus cases. Therefore, the influence of the elastic modulus is relatively small.  All numerical results remain in good agreement with the analytical solution.
\begin{figure}[h!]
       \centering
	   \begin{subfigure}{0.48\linewidth}
		\includegraphics[scale=1]{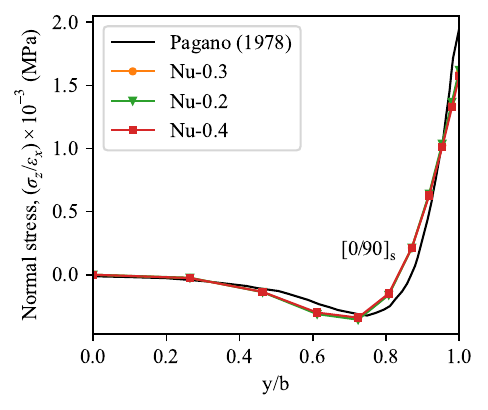}
		\caption{$\sigma_z$ at the central plane}
		\label{fig:Cross-ply-Nu-mid}
	   \end{subfigure}
    ~
	     \begin{subfigure}{0.48\linewidth}
		 \includegraphics[scale=1]{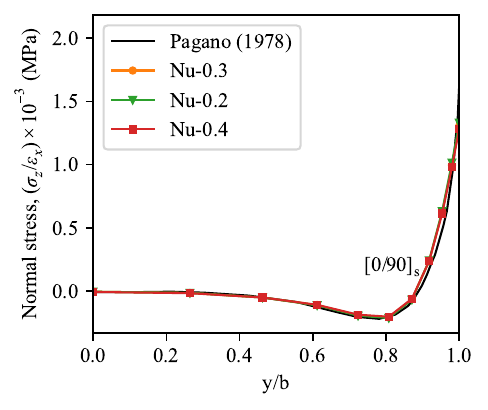}
		 \caption{$\sigma_z$ at the interface between $0^{\circ}$ and $90^{\circ}$}
		 \label{fig:Cross-ply-Nu-H-sigmaZ}
	      \end{subfigure}
      \vfill  
      \begin{subfigure}{0.48\linewidth}
		 \includegraphics[scale=1]{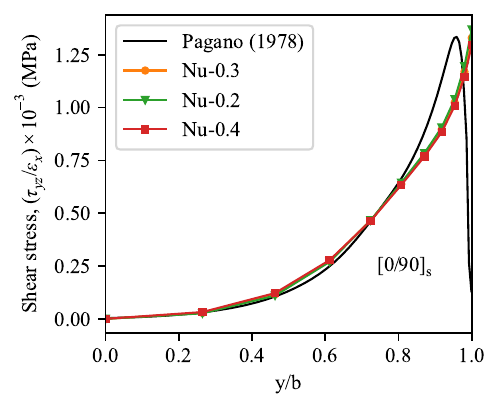}
		 \caption{$\tau_{yz}$ at the interface between $0^{\circ}$ and $90^{\circ}$}
		 \label{fig:Cross-ply-Nu-H-tauYZ}
	      \end{subfigure}
	\caption{Distributions of interlaminar normal and shear stresses with different Poisson's ratios}
	\label{fig:Cross-ply-Nu}
\end{figure}

Figure~\ref{fig:Cross-ply-Nu} presents the interlaminar stress distributions obtained for different Poisson's ratios. Compared with Figure~\ref{fig:Cross-ply-E}, the influence of Poisson's ratio is even smaller. No significant differences are observed among the different cases, either at the laminate mid-plane or at the $(0^\circ/90^\circ)$ interface. These results indicate that the influence of Poisson's ratio on the predicted stress distributions is negligible. This behavior can be attributed to the fact that Poisson's ratio affects only the shear modulus, while the normal penalty stiffness, $K_n$, remains unchanged. Therefore, the variation in the interlaminar normal stress ($\sigma_z$) is negligible. As for the shear penalty stiffness, the variations are also relatively small. Therefore, the predicted shear stress, $\tau_{yz}$, is insensitive to Poisson's ratio.

In summary, the predicted stress distributions exhibit only minor variations over the range of epoxy resin properties considered. The predicted stress distributions remain in close agreement with the analytical solution. For consistency with the previous validation cases, the resin-rich layer properties are taken as $E=3000\,\text{MPa}$ and $\nu=0.3$ in all subsequent analyses.

\subsubsection{Mesh convergence analysis of the solid-element model}
In this section, a mesh convergence study of the solid-element models is presented. The study has two objectives. The first is to compare the predictions of the solid-element model, the proposed structural-element model, and the available analytical solutions. This comparison provides a reference for assessing the accuracy of the solid-element model. The second is to establish a reliable benchmark for problems without analytical solutions. This is particularly important for predicting through-thickness stress distributions in complex laminate lay-ups, such as the general laminate configurations presented in Section~\ref{sec:General-lay-up}. In these cases, the proposed formulation is validated through quantitative comparisons with high-fidelity solid-element simulations.

\begin{figure}[h!]
      \centering
	   \begin{subfigure}{0.3\linewidth}
		\includegraphics[width=\linewidth]{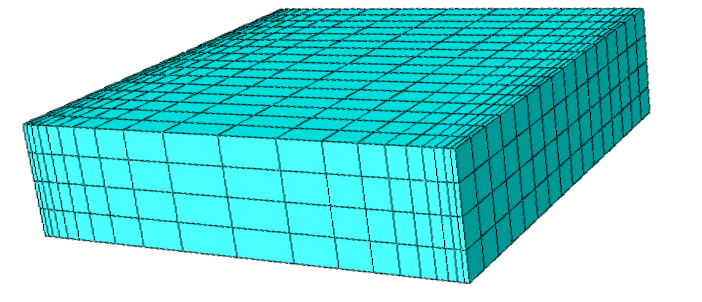}
		\caption{Mesh size: 1 element per layer}
		\label{fig:solid_mesh-subfig1}
	   \end{subfigure}
    ~
	   \begin{subfigure}{0.3\linewidth}
		\includegraphics[width=\linewidth]{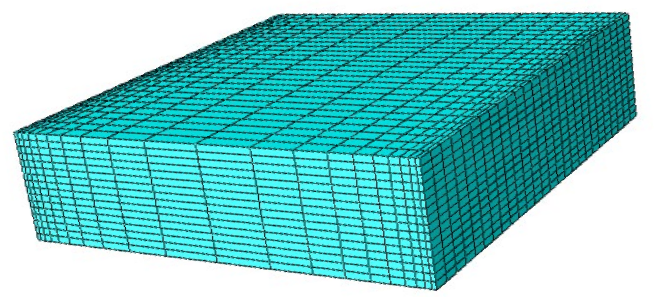}
		\caption{Mesh size: 4 elements per layer}
		\label{fig:solid_mesh-subfig2}
	    \end{subfigure} 
    ~
      \begin{subfigure}{0.3\linewidth}
		\includegraphics[width=\linewidth]{figures/Figures-SolidMeshT4.pdf}
		\caption{Mesh size: 8 elements per layer}
		\label{fig:solid_mesh-subfig3}
        \end{subfigure} 
    \caption{Mesh sizes of solid-element model} 
	\label{fig:solid-model mesh}
\end{figure}

The mesh convergence study focuses only on mesh refinement in the through-thickness direction. The mesh density along the laminate width is identical to that used in the structural-element model. Three through-thickness mesh schemes are considered, as illustrated in Figure~\ref{fig:solid-model mesh}: (1) one solid element per ply (four elements through the thickness); (2) four solid elements per ply; and (3) eight solid elements per ply.

\begin{figure}[h!]
      \centering
	   \begin{subfigure}{0.48\linewidth}
		\includegraphics[scale=1]
        {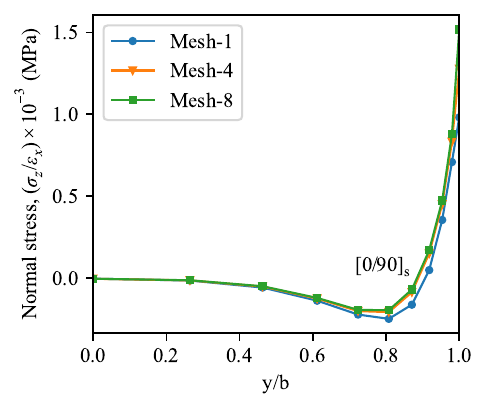}
		\caption{Interlaminar normal stress $\sigma_z$}
		\label{fig:sigmaZ_solid_h}
	   \end{subfigure}
	   \begin{subfigure}{0.48\linewidth}
		\includegraphics[scale=1]
        {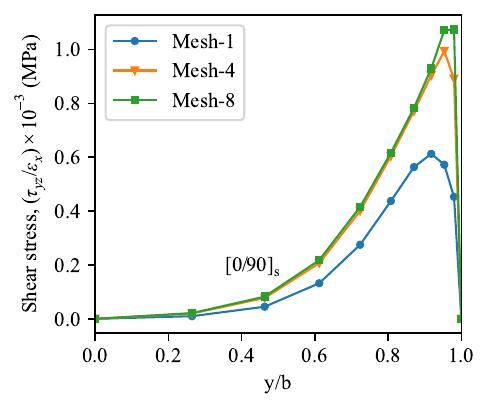}
		\caption{Interlaminar shear stress $\tau_{yz}$}
		\label{fig:tauYZ_solid_h}
	    \end{subfigure}
	\caption{{Interlaminar stress distributions at the interface between $0^{\circ}$ and $90^{\circ}$ }}
	\label{fig:Solid-comparison}
\end{figure}

Figure~\ref{fig:Solid-comparison} presents the stress distributions at $(0^{\circ}/90^{\circ})$ interface for different through-thickness mesh densities. The results show that employing only one solid element per ply does not provide sufficient accuracy. In contrast, refining the mesh to four or eight elements per ply results in a clear improvement. This improvement is particularly evident for the interlaminar shear stress, $\tau_{yz}$, shown in Figure~\ref{fig:tauYZ_solid_h}. Based on these results, all solid-element models used in this study employ eight elements per ply through the thickness to ensure accurate stress predictions.

\subsubsection{Comparison with analytical and solid-element solutions}
\begin{figure}[h!]
       \centering
	   \begin{subfigure}{0.48\linewidth}
		\includegraphics[scale=1]{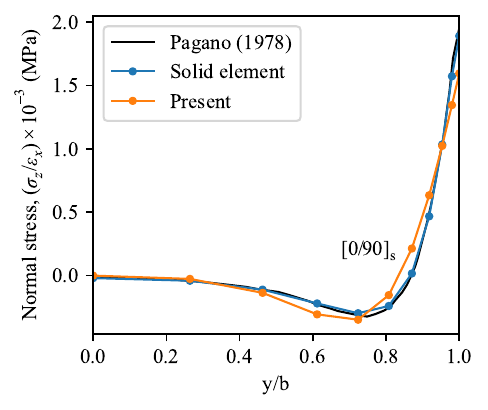}
		\caption{$\sigma_z$ at the central plane}
		\label{fig:Cross-ply-mid-compare}
	   \end{subfigure}
    ~
	     \begin{subfigure}{0.48\linewidth}
		 \includegraphics[scale=1]{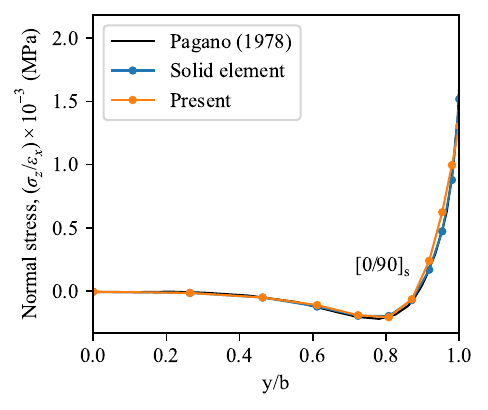}
		 \caption{$\sigma_z$ at the interface between $0^{\circ}$ and $90^{\circ}$}
		 \label{fig:Cross-ply-H-sigmaZ-compare}
	      \end{subfigure}
      \vfill  
      \begin{subfigure}{0.48\linewidth}
		 \includegraphics[scale=1]{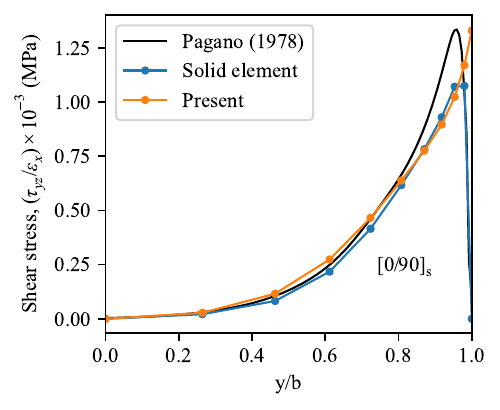}
		 \caption{$\tau_{yz}$ at the interface between $0^{\circ}$ and $90^{\circ}$}
		 \label{fig:Cross-ply-tauYZ-compare}
	      \end{subfigure}
	\caption{Results comparison for the distributions of interlaminar normal and shear stresses}
	\label{fig:Cross-ply-compare}
\end{figure}

The previous validation results are summarized through direct comparisons of the analytical solutions, the proposed structural-element results, and the solid-element results. The results are presented in Figures~\ref{fig:Cross-ply-mid-compare}, \ref{fig:Cross-ply-H-sigmaZ-compare}, and~\ref{fig:Cross-ply-tauYZ-compare}. The comparisons show that both numerical models accurately reproduce the analytical stress distributions.

\subsection{Results for angle-ply laminates $[+\theta/-\theta]_{s}$}
\label{subsec-angle-ply-theta}
In this section, the stress distributions in the $[+\theta/-\theta]_s$ angle-ply laminate are investigated. The geometric dimensions and boundary conditions are identical to those of the cross-ply laminate presented in Section~\ref{subsec-cross-ply}. Different composite material properties are adopted. The corresponding material properties are listed in Table~\ref{tab:Angleply_Material}.
\begin{table}[h!]
  \centering
  \caption{Material properties of angle-ply laminate \cite{mittelstedt2007pipes}}
    \begin{tabular}{lll}
    \toprule
 \multicolumn{3}{l}{\textbf{Material properties of $[+\theta/-\theta]_s$ laminate} }   \\
 \midrule
    $E_{11}$= 13200 MPa & $E_{22}$= 10800 MPa & $E_{33}$ = 10800 MPa \\[0.3em]
    $\nu_{12}$ = 0.238 & $\nu_{13}$ = 0.238 & $\nu_{23}$ = 0.49  \\[0.3em]
    $G_{12}$ = 5650 MPa & $G_{13}$ = 5650 MPa & $G_{23}$ = 3360 MPa \\[0.5em]
    \bottomrule
    \end{tabular}%
  \label{tab:Angleply_Material}%
\end{table}%

\begin{figure}[h!]
	\centering
	\includegraphics[scale=1]{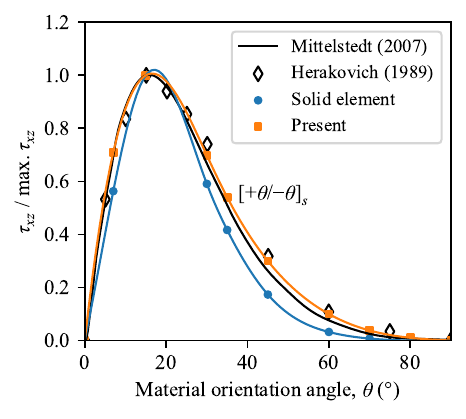}
	\caption{Interlaminar shear stress $\tau_{xz}$ at the $+\theta/-\theta$ interface $(z = h_0)$}
	\label{fig:theta/-theta}
\end{figure}

Unlike the cross-ply laminate, the interlaminar stresses $\tau_{yz}$ and $\sigma_z$ in the $[+\theta/-\theta]_s$ laminate are negligible, as discussed in Subsection~\ref{subsec-cross-ply}. In contrast, previous studies~\cite{pipes1970interlaminar,pipes1974interlaminar,kassapoglou1986efficient,kassapoglou1987closed} have shown that the interlaminar shear stress, $\tau_{xz}$, becomes significant near the free edge of the $[+45^\circ/-45^\circ]$ interface. Therefore, the stress distributions are normalized with respect to the corresponding peak values for each lay-up angle. The normalized results are then compared with the analytical solution~\cite{mittelstedt2007pipes} and the finite element results reported by \citet{herakovich1989free}. The corresponding comparison results are presented in Figure~\ref{fig:theta/-theta}.

Figure~\ref{fig:theta/-theta} shows that the normalized interlaminar shear stress, $\tau_{xz}/\max(\tau_{xz})$, predicted by the proposed method. The result is in excellent agreement with both the analytical solutions and the finite element results. Combined with the validation results for the cross-ply laminates, these results further confirm the accuracy of the proposed formulation. Therefore, the proposed method provides a computationally efficient alternative to solid-element models for predicting interlaminar stress distributions.

\subsection{Results for quasi-isotropic laminates $[+45^{\circ}/0^{\circ}/-45^{\circ}/90^{\circ}]_{s}$}
\label{subsec-Reddy}
\begin{figure}[h!]
	\centering
	\includegraphics[width=0.6\linewidth]{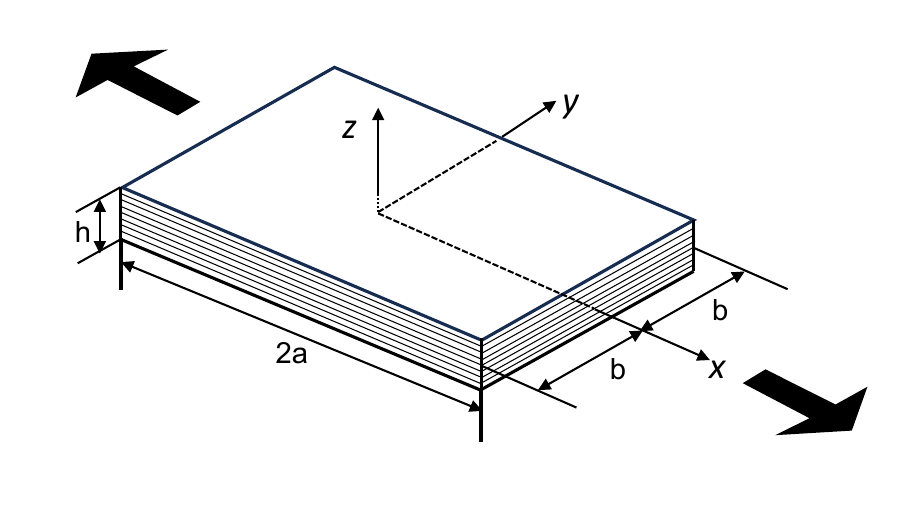}
	\caption{Quasi-isotropic laminate geometry}
	\label{fig:UAE-Reddy}
\end{figure}

In Sections~\ref{subsec-cross-ply} and~\ref{subsec-angle-ply-theta}, the proposed method was validated using four-ply laminates with relatively simple stacking sequences. In this section, the method is further validated using the eight-ply quasi-isotropic laminate proposed by \citet{reddy2004mechanics}. The predicted stress distributions are compared with the corresponding analytical solutions. It should be noted that the geometry of this model, shown in Figure~\ref{fig:UAE-Reddy}, differs from that described in Section~\ref{subsec:UAE model}. The specific geometric parameters are summarized below:
\begin{align}\label{eq:Reddy-geo}
    (45/0/-45/90)_s \,\textrm{laminate}: h=8h_0, \,b= 60h_0,\, a=10b 
\end{align}

In the present study, the thickness of each ply is set to $h_0=0.125\,\mathrm{mm}$. The remaining geometric dimensions of the laminate are determined from Equation~\ref{eq:Reddy-geo}, and the corresponding values are listed in Table~\ref{tab:Reddy-geo}. In addition, the material properties are summarized in Table~\ref{tab:Reddy_Material}. To achieve the tensile strain of $\varepsilon_x=0.01$ adopted by \citet{reddy2004mechanics}, a tensile displacement of $0.75 \,\mathrm{mm}$ is applied at both ends of the laminate.
\begin{table}[h!]
  \centering
  \caption{Geometric parameters of the quasi-isotropic laminate benchmark model}
    \begin{tabular}{llllll}
    \toprule
    Parameter &  $h_0$ & $h$  & $b$ (width) & $a$ (length)  \\[0.3em]
 \midrule
    value (mm) & 0.125 & 1 & 7.5 & 75  \\[0.3em]
    \bottomrule
    \end{tabular}%
  \label{tab:Reddy-geo}%
\end{table}

\begin{table}[h!]
  \centering
  \caption{Material properties of the quasi-isotropic laminate~\cite{reddy2004mechanics}}
    \begin{tabular}{lll}
    \toprule
 \multicolumn{3}{l}{\textbf{Material properties of $[+45/0/-45/90]_s$ laminate}}   \\
 \midrule
    $E_{11}$= 134448 MPa & $E_{22}$= 10204 MPa & $E_{33}$ = 10204 MPa \\[0.3em]
    $\nu_{12}$ = 0.3 & $\nu_{13}$ = 0.3 & $\nu_{23}$ = 0.3  \\[0.3em]
    $G_{12}$ = 5516 MPa & $G_{13}$ = 5516 MPa & $G_{23}$ = 5516 MPa \\[0.5em]
    \bottomrule
    \end{tabular}%
  \label{tab:Reddy_Material}%
\end{table}%

\begin{figure}[h!]
      \centering
	   \begin{subfigure}{0.48\linewidth}
		\includegraphics[scale=1]{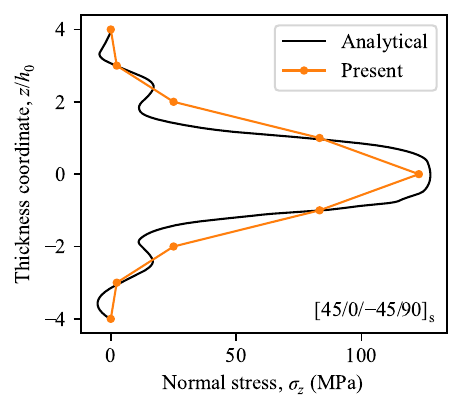}
		\caption{Interlaminar normal stress $\sigma_z$}
		\label{fig:sigmaZ_reddy}
	   \end{subfigure}
	   \begin{subfigure}{0.48\linewidth}
		\includegraphics[scale=1]
        {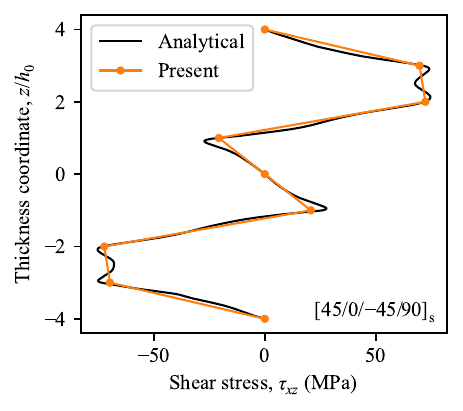}
		\caption{Interlaminar shear stress $\tau_{xz}$}
		\label{fig:tauXZ_reddy}
	    \end{subfigure}
	\caption{{Interlaminar stress distributions through the thickness near the free edge ($y=0.998b$) in $[45/0/-45/90]_s$ laminate under axial extension}}
	\label{fig:Reddy-comparison}
\end{figure}

A comparison between the interlaminar stresses ($\sigma_z$ and $\tau_{xz}$) predicted by the proposed method and those obtained from Mittelstedt's analytical solution~\cite{mittelstedt2007pipes} is shown in Figure~\ref{fig:Reddy-comparison}. The comparison is performed through the laminate thickness at the location ($x=-0.0115a, y=0.998b$). As observed, the results obtained using the proposed numerical method are in good agreement with those predicted by Mittelstedt's layerwise analytical method. The orange markers in Figure~\ref{fig:Reddy-comparison} represent the stress values at the ply interfaces. All interface stresses lie on or very close to the analytical solution. The orange solid line is obtained by applying the linear interpolation procedure described in Section~\ref{subsec-Interpolation along thickness}. As shown in the right panel, the predicted distribution of the interlaminar shear stress ($\tau_{xz}$) agrees well with the analytical solution. In contrast, small discrepancies are observed in the normal stress $(\sigma_z)$. These discrepancies arise because the stress variation between several adjacent interfaces is non-monotonic and cannot be accurately captured by linear interpolation. Employing a higher-order interpolation scheme is expected to further improve the agreement with the analytical solution.

A limitation of the proposed method is observed at the $[0^\circ/-45^\circ]$ interface, where the predicted $\sigma_z$ values are slightly higher than the analytical solution. This discrepancy is attributed to the use of a single shell element to represent each ply, without additional integration points through the thickness. Complex through-thickness stress variations cannot be captured exactly. Nevertheless, the overall agreement with the analytical solution remains good. These results demonstrate the reliability of the proposed method. They also confirm its applicability to thick laminates containing four or more plies, which are commonly considered in free-edge stress analyses of composite laminates.

\subsection{Results for laminates with general lay-ups}
\label{sec:General-lay-up}
The previous validation cases focused on laminate configurations with available analytical solutions. In practical engineering applications, however, composite laminates often employ complex or arbitrary stacking sequences for which no analytical solutions are available. Therefore, the proposed method is further validated using two arbitrary laminate lay-ups inspired by \citet{andakhshideh2013interlaminar}. The two stacking sequences considered are: (a) $[0^{\circ}/+55^{\circ}/+35^{\circ}/-35^{\circ}]_s$ and (b) $[0^{\circ}/+30^{\circ}/-45^{\circ}/+45^{\circ}]_s$. Since both laminates contain eight plies, the same geometric dimensions, material properties, and loading conditions as those described in Section~\ref{subsec-Reddy} are adopted.

\subsubsection{$[0^{\circ}/+55^{\circ}/+35^{\circ}/-35^{\circ}]_s$  lay-up laminate}

\begin{figure}[h!]
       \centering
	   \begin{subfigure}{0.48\linewidth}
		\includegraphics[scale=1]{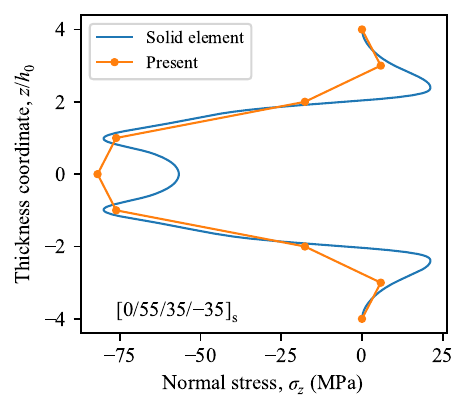}
		\caption{Normal stress $\sigma_{z}$}
		\label{fig:General_SigmaZ_thickness_layup1}
	   \end{subfigure}
    ~
	     \begin{subfigure}{0.48\linewidth}
		 \includegraphics[scale=1]{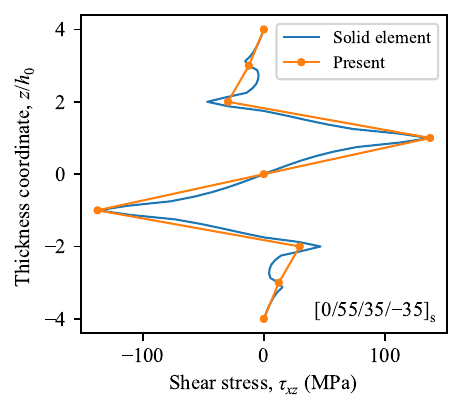}
		 \caption{Shear stress $\tau_{xz}$}
		 \label{fig:General_TauXZ_thickness_layup1}
	      \end{subfigure}
	\caption{Through-thickness distributions of interlaminar stresses in the $[0^{\circ}/+55^{\circ}/+35^{\circ}/-35^{\circ}]_s$ laminate $(y=0.998b)$ under axial extension} 
	\label{fig:General_thickness_layup1}
\end{figure}

Figure~\ref{fig:General_thickness_layup1} compares the through-thickness distributions of the interlaminar normal stress ($\sigma_z$) and the interlaminar shear stress ($\tau_{xz}$) for the $[0/55/35/-35]_s$ laminate. The predicted $\sigma_z$ distribution agrees closely with the solid-element solution. Only small discrepancies are observed in the mid-plane. Therefore, the stress distribution cannot be fully captured by the linear interpolation procedure. Figure~\ref{fig:General_TauXZ_thickness_layup1} shows the corresponding interlaminar shear stress distribution. Excellent agreement is obtained throughout the laminate thickness, including at the ply interfaces represented by the orange markers. In conclusion, the proposed method provides predictions of the interlaminar stress distributions that are comparable in accuracy to those obtained using the solid-element model.

\begin{figure}[h!]
       \centering
	   \begin{subfigure}{0.48\linewidth}
		\includegraphics[scale=1]{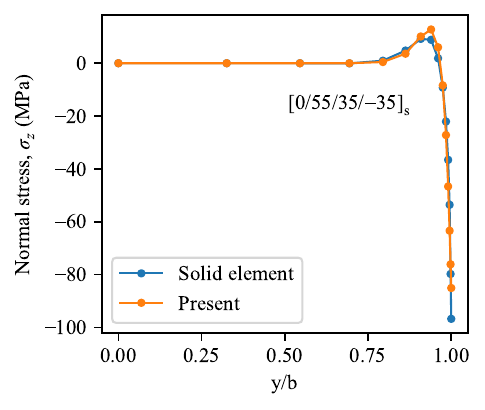}
		\caption{Normal stress $\sigma_{z}$}
		\label{fig:General_SigmaZ_WH}
	   \end{subfigure}
    ~
	     \begin{subfigure}{0.48\linewidth}
		 \includegraphics[scale=1]{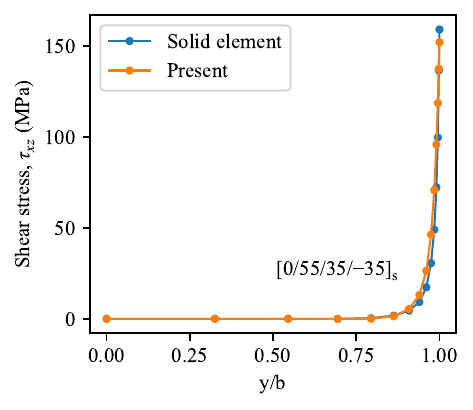}
		 \caption{Shear stress $\tau_{xz}$}
		 \label{fig:General_TauXZ_WH}
	      \end{subfigure}
      \vfill  
      \begin{subfigure}{0.48\linewidth}
		 \includegraphics[scale=1]{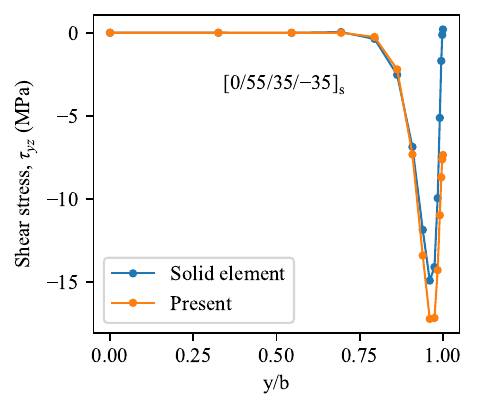}
		 \caption{Shear stress $\tau_{yz}$}
		 \label{fig:General_TauYZ_WH}
	      \end{subfigure}
	\caption{Interlaminar stress distributions at the interface between $35^{\circ}$ and $-35^{\circ}$ }
	\label{fig:General_WH}
\end{figure}

Following the through-thickness comparison shown in Figure~\ref{fig:General_thickness_layup1}, the interlaminar stress distributions along the laminate width are compared at the interface between the $35^\circ$ and $-35^\circ$ plies, as shown in Figure~\ref{fig:General_WH}. The proposed method shows excellent agreement with the solid-element results over the laminate width. In particular, the free-edge stress singularity is accurately captured. A small discrepancy is observed in the predicted $\tau_{yz}$ distribution, although the overall trend remains consistent with the solid-element solution. This discrepancy is attributed to the interpolation procedure rather than the proposed formulation itself.

\subsubsection{$[0^{\circ}/+30^{\circ}/-45^{\circ}/+45^{\circ}]_s$ lay-up laminate}

\begin{figure}[h!]
       \centering
	   \begin{subfigure}{0.48\linewidth}
		\includegraphics[scale=1]{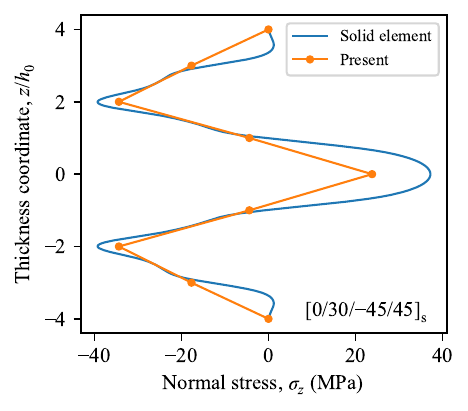}
		\caption{Normal stress $\sigma_{z}$}
		\label{fig:General_SigmaZ_thickness_layup2}
	   \end{subfigure}
    ~
	     \begin{subfigure}{0.48\linewidth}
		 \includegraphics[scale=1]{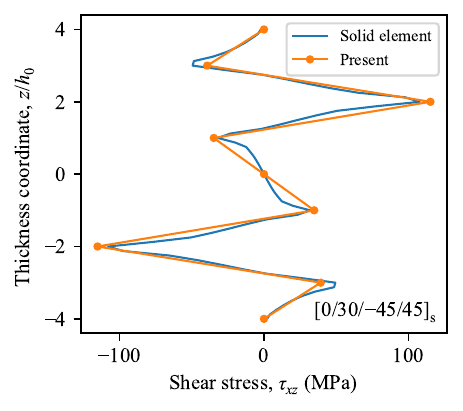}
		 \caption{Shear stress $\tau_{xz}$}
		 \label{fig:General_TauXZ_thickness_layup2}
	      \end{subfigure}
	\caption{Through-thickness distributions of interlaminar stresses in the  $[0^{\circ}/+30^{\circ}/-45^{\circ}/+45^{\circ}]_s$ laminate $(y=0.998b)$ under axial extension} 
	\label{fig:General_thickness_layup2}
\end{figure}

After validating the $[0/55/35/-35]_s$ laminate, the $[0/30/-45/45]_s$ laminate is considered to further assess the general applicability of the proposed method. Compared with the previous lay-up, this result shows even better agreement with the solid-element solution because the through-thickness stress distribution is nearly monotonic. The proposed method accurately reproduces the maximum and minimum stresses, as well as the intermediate stress variation. A small discrepancy is observed only for $\sigma_z$ in the mid-plane, where the predicted values are slightly lower than those of the solid-element model. This difference is attributed to the finer through-thickness discretization of the solid-element model, which captures the local stress singularity more accurately.

\begin{figure}[h!]
       \centering
	   \begin{subfigure}{0.48\linewidth}
		\includegraphics[scale=1]{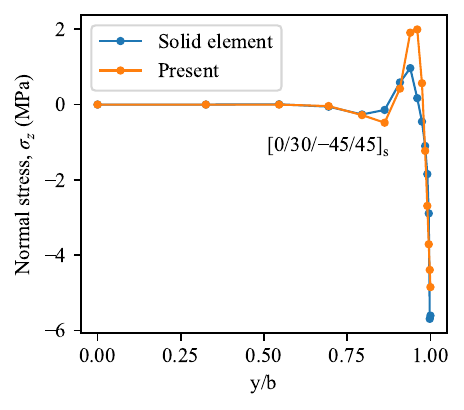}
		\caption{Normal stress $\sigma_{z}$}
		\label{fig:General_SigmaZ_WH_layup2}
	   \end{subfigure}
    ~
	     \begin{subfigure}{0.48\linewidth}
		 \includegraphics[scale=1]{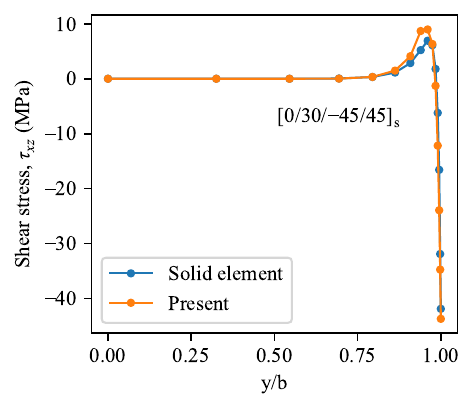}
		 \caption{Shear stress $\tau_{xz}$}
		 \label{fig:General_TauXZ_WH_layup2}
	      \end{subfigure}
      \vfill  
      \begin{subfigure}{0.48\linewidth}
		 \includegraphics[scale=1]{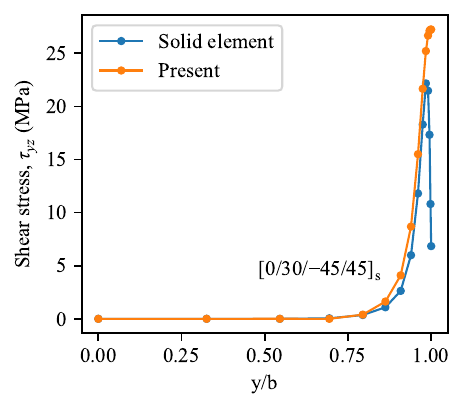}
		 \caption{Shear stress $\tau_{yz}$}
		 \label{fig:General_TauYZ_WH_layup2}
	      \end{subfigure}
	\caption{Interlaminar stress distributions along the width direction at the interface between the $-45^{\circ}$ and $45^{\circ}$ plies of $[0^{\circ}/+30^{\circ}/-45^{\circ}/+45^{\circ}]_s$ laminate}
	\label{fig:General_WH_layup2}
\end{figure}

Similarly, the interlaminar stress distributions along the laminate width for the $[0/30/\minus45/45]_s$ laminate are presented in Figure~\ref{fig:General_WH_layup2}. The predicted normal stress $\sigma_z$ follows the same trend as the solid-element solution. It begins with a small compressive stress, reaches a tensile peak at approximately $y/b=0.94$, and then decreases rapidly toward the free edge. The only difference is a slightly higher peak tensile stress predicted by the proposed method. This discrepancy is attributed to the relatively low magnitude of the normal stress in this ply. It makes the prediction more sensitive to numerical interpolation. The interlaminar shear stress distributions are shown in Figures~\ref{fig:General_TauXZ_WH_layup2} and~\ref{fig:General_TauYZ_WH_layup2}. The predicted shear stress $\tau_{xz}$ is in excellent agreement with the solid-element results. The proposed method accurately captures both the stress extrema and the free-edge stress singularity. For the remaining shear stress component ($\tau_{yz}$), a small discrepancy is observed only at the free edge. This behavior is consistent with the interpolation-related limitation discussed previously.

\subsubsection{Normal stress distribution at the laminate mid-plane}
\begin{figure}[h!]
	\centering
	\begin{subfigure}{0.48\linewidth}
	\includegraphics[scale=1]{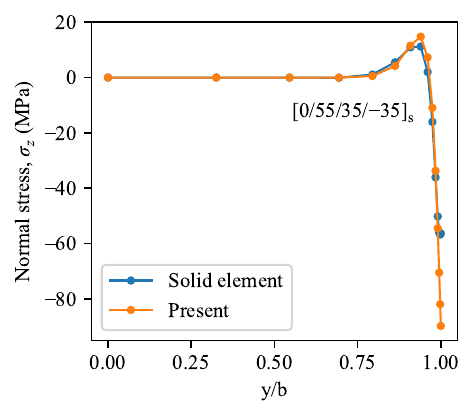}
    \caption{$[0^{\circ}/+55^{\circ}/+35^{\circ}/-35^{\circ}]_s$ laminate}
	\label{fig:General_SigmaZ_mid_layup1}
	\end{subfigure}
    ~
	\begin{subfigure}{0.48\linewidth}
	  \includegraphics[scale=1]{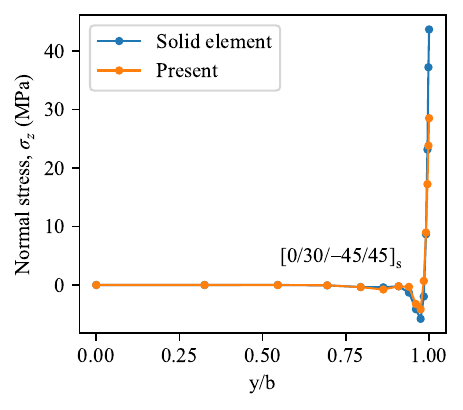}
    \caption{$[0^{\circ}/+30^{\circ}/-45^{\circ}/+45^{\circ}]_s$ laminate}
	\label{fig:General_SigmaZ_mid_layup2}
	\end{subfigure}
	\caption{Interlaminar normal stress $\sigma_z$ at the mid-plane}
	\label{fig:General_SigmaZ_mid}
\end{figure}

As in the previous sections, the stress distribution along the laminate width is also evaluated at the mid-plane. The corresponding results are presented in Figure~\ref{fig:General_SigmaZ_mid}. Since the interlaminar shear stresses vanish throughout the mid-plane, only the normal stress ($\sigma_z$) is considered. As shown in the figure, the proposed method shows excellent agreement with the solid-element results. It accurately reproduces both the overall stress distribution and the corresponding peak values.

\subsubsection{Displacement distribution along the thickness}
\begin{figure}[h!]
      \centering
	   \begin{subfigure}{0.48\linewidth}
		\includegraphics[scale=1]{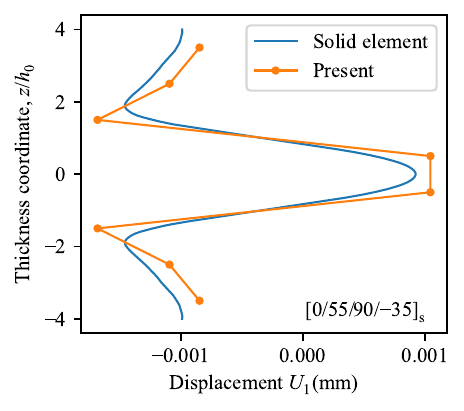}
		\caption{$[0^{\circ}/+55^{\circ}/+35^{\circ}/-35^{\circ}]_s$ laminate}
		\label{fig:eneral_8Layers_U1}
	   \end{subfigure}
	   \begin{subfigure}{0.48\linewidth}
		\includegraphics[scale=1]{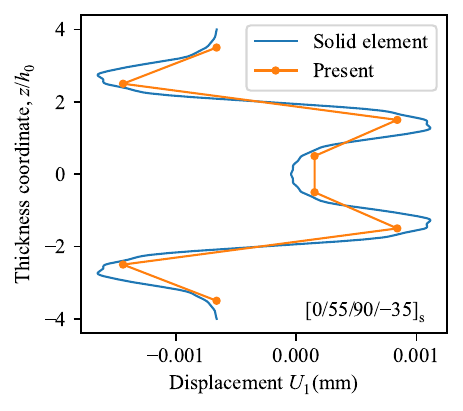}
		\caption{$[0^{\circ}/+30^{\circ}/-45^{\circ}/+45^{\circ}]_s$ laminate}
		\label{fig:General_8Layers_U2}
	    \end{subfigure}
	\caption{Displacement through the thickness}
	\label{fig:General_8Layers_disp}
\end{figure}

Finally, the through-thickness displacement distribution is validated, as shown in Figure~\ref{fig:General_8Layers_disp}. Despite the highly nonlinear through-thickness displacement distribution, the proposed method accurately captures both the overall variation and the displacement values at the ply interfaces. Small discrepancies are observed only in regions where the displacement varies rapidly. These discrepancies are attributed to the linear interpolation procedure. These results further demonstrate the accuracy of the proposed method in recovering the through-thickness displacement field for laminates with general stacking sequences.

\subsubsection{Computational performance}
The proposed method employs shell elements, with each ply represented by a single element layer. Therefore, the computational cost is significantly lower than that of the corresponding solid-element model. The CPU times are compared in Table~\ref{tab:Computational performances}. The proposed method reduces the CPU time per increment by approximately 45\% while maintaining accurate predictions of the interlaminar stress distributions.
\begin{table}[h]
	\centering
	\caption{Comparison of CPU time per increment (unit: second)}
	\begin{tabular}{ l l}
		\toprule
         General lay-up laminate& CPU time per increment \\[0.3em]
         \midrule
         Solid model  & 9.52  \\[0.3em]
         \midrule
         Present model & 5.21 \\[0.3em]
         \midrule
         Reduction by proposed method & 45.3\%  \\[0.3em]
         \bottomrule 
	\end{tabular}
	\label{tab:Computational performances}
\end{table}

\section{Summary and conclusions}
\label{sec:summary}
This work proposes a resin-rich layer-based formulation for the cohesive penalty stiffness and demonstrates its application to the recovery of three-dimensional stress fields in shell-element models of composite laminates. The proposed penalty stiffness is derived from the material properties and thickness of the resin-rich layer. It enables accurate simulation of uniform axial extension and free-edge stresses using the structural cohesive elements proposed by \citet{ai2025structural}. The proposed formulation preserves the Kirchhoff–Love kinematic assumptions while providing a physically based definition of the cohesive penalty stiffness. In addition, the proposed method shows low sensitivity to variations in the properties of the resin-rich layer and is applicable to a wide range of composite laminates. It also provides a physically based foundation for the further development of structural cohesive element models.

The proposed method has been validated using cross-ply, angle-ply, quasi-isotropic, and general symmetric laminate configurations. Excellent agreement is obtained with both analytical solutions and solid-element models. Compared with solid-element models, the proposed method requires a much coarser through-thickness discretization while maintaining comparable accuracy in predicting interlaminar stress distributions. These results confirm the applicability of the proposed method to composite laminates with complex stacking sequences.

Finally, the proposed method improves previous work on Kirchhoff-Love ply and cohesive elements for composites modelling, where accurate out-of-plane stresses can now be obtained throughout the thickness of the laminate. It establishes a framework for simulating combined in-plane damage and delamination efficiently on coarse meshes in the future.

\section*{Acknowledgements}
The first author would like to acknowledge funding support from the China Scholarship Council (No.201906290034) for this research.


\bibliographystyle{elsarticle-num-names} 
\bibliography{cas-refs}







\end{document}